\documentclass[pra,showpacs,twocolumn,floatfix,superscriptaddress]{revtex4}

\usepackage{graphicx}
\usepackage{dcolumn}
\usepackage{bm}
\usepackage{amsmath}

\usepackage{color}
\usepackage[normalem]{ulem}
\usepackage{braket}

\newcommand{\be}{\begin{equation}}
\newcommand{\ee}{\end{equation}}
\newcommand{\ba}{\begin{eqnarray}}
\newcommand{\ea}{\end{eqnarray}}
\newcommand{\nn}{\nonumber \\}
\def\ket#1{| #1 \rangle}
\def\bra#1{\langle #1 |}

\begin{document}

\title{Theory of open quantum dynamics with hybrid noise}

\author{Anatoly~Yu.~Smirnov}
\affiliation{D-Wave Systems Inc., 3033 Beta Avenue, Burnaby BC}
\author{Mohammad~H.~Amin}
\affiliation{D-Wave Systems Inc., 3033 Beta Avenue, Burnaby BC}
\affiliation{Department of Physics, Simon Fraser University,
Burnaby, British Columbia, Canada V5A 1S6}

\date{\today}

\begin{abstract}
We develop a theory to describe dynamics of a non-stationary open
quantum system interacting with a hybrid environment, which includes
high-frequency and low-frequency noise components. One part of the
system-bath interaction is treated in a perturbative manner, whereas
the other part is considered exactly. This approach allows us to
derive a set of master equations where the relaxation rates are
expressed as convolutions of the Bloch-Redfield and Marcus formulas.
Our theory enables analysis of systems that have extremely small
energy gaps in the presence of a realistic environment. As an
illustration, we apply the theory to the 16-qubit quantum annealing
problem with dangling qubits and show good agreement with
experimental results.
\end{abstract}


 \maketitle

\section{Introduction}

The theory of open quantum dynamics
\cite{Leggett87,Slichter90,Walls94} is an important and active area
of physics with applications in nanotechnology, chemical physics,
quantum biology, and quantum information. In open quantum theories,
the system under consideration is assumed to interact with an
environment that has many degrees of freedom. Because details of the
environmental Hamiltonian are usually unknown, measurable quantities
such as temperature $T$ or noise spectrum $S(\omega)$ are used to
describe the average statistical behavior of the environment. An
open quantum model, therefore, provides a set of differential
equations that describe the statistical dynamics of the quantum
system, taking the temperature and the spectrum of the bath as input
parameters.

Increased research and development of technology in quantum
computing is renewing interest in open quantum modeling. One such
promising computation scheme is quantum annealing (QA)
\cite{Kadowaki98,Santoro02,Johnson11}  (in particular, adiabatic
quantum computation \cite{Farhi01}). In QA, the system is evolved
slowly so that it stays at or near the ground state throughout the
evolution. At the end of the evolution, the system will occupy a
low-energy state of the final Hamiltonian, which may represent a
solution to an optimization or a sampling problem.

Open quantum dynamics of a QA processor have been studied
theoretically \cite{Sahel06,AminLove,Albash12,Albash15}. These
models assume weak coupling to an environment, which is typically
taken to have Ohmic spectrum with large high-frequency content. This
limit is well described by the Bloch-Redfield theory
\cite{Leggett87,Slichter90,Walls94,AminLove,ES81}. Realistic qubits
\cite{Harris10}, however, suffer from strong interaction with
low-frequency noise (in particular, noise with 1/f-like spectrum).
Incoherent dynamics of a qubit coupled to such an environment are
described by the Marcus theory \cite{Marcus93,Cherepanov01,LNT17}. A
complete ({\em hybrid}) open quantum model should account for both
low-frequency and high-frequency environments. Such a model for a
single qubit has been developed and agreement with experiment has
been demonstrated \cite{Amin08,AminBrito09,Lanting11}. A
generalization of this theory to multiqubit systems has also been
developed and compared with experimental observation
\cite{Boixo16,Boixo14}. Several attempts to combine the
Bloch-Redfield and Marcus methods have been undertaken in chemical
physics and quantum biology(see, for example,
Refs.\cite{Yang02,Ghosh11,Lambert13}).

In this paper, we expand the work of Refs.~\cite{Boixo16,Boixo14}.
We provide a systematic and detailed derivation of a hybrid open
quantum model, which agrees with the results of
Ref.~\cite{Boixo16,Boixo14} for problems with large spectral gaps.
Our theory, however, can also be applied to small-gap problems with
nonstationary Hamiltonians, for which the model in
\cite{Boixo16,Boixo14} is not applicable. We provide an intuitively
appealing and computationally convenient form for the transition
rates in terms of a convolution between Redfield and Marcus
formulas. As an example, we investigate a dissipative evolution of a
16-qubit system strongly interacting with low-frequency noise and
weakly coupled to a high-frequency environment. The problem is
characterized by an extremely small gap in the energy spectrum of
qubits in the middle of annealing. Solving the problem requires the
right combination of Bloch-Redfield and Marcus approaches as well as
a proper consideration of the calculation basis, which takes into
account the nonstationary effects. The results of the present paper
can also be applied to any other open quantum system.

The paper is organized in the following way. Section~\ref{SingleQ}
describes a single-qubit system to provide the necessary intuition
before moving to more complicated multi-qubit problems.
Section~\ref{HamSec} formulates the Hamiltonian and introduces
important definitions and notations for the system and the bath.
Master equations for the probability distribution of the quantum
system are derived in Section~\ref{IntPic}. Section~\ref{BRM}
presents the relaxation rates as convolution integrals of the
Bloch-Redfield and Marcus envelopes. In the Section \ref{Limits} we
show that in equilibrium the master equations obey the detailed
balance conditions. We also demonstrate that the convolution
expression for the relaxation rates turns into the Marcus or to
Bloch-Redfield formulas in the corresponding limits. Dissipative
dynamics of a 16-qubit system with an extremely small energy gap is
considered in Section~\ref{Dickson16}. A brief compilation of the
commonly encountered notations is presented in Appendix
\ref{ApxNot}. In other Appendixes we provide a detailed derivation
of many important formulas.

\section{Single-qubit system}
\label{SingleQ}

We begin with a single-qubit system that has a Hamiltonian
\ba \label{HamTLS} H_S = - \frac{\Delta}{2} \, \sigma_x -
\frac{h}{2}\, \sigma_z, \ea
with the Pauli matrices $\sigma_x, \sigma_y, \sigma_z$, tunneling
amplitude $\Delta$, and bias $h$. The ground state, $\ket{1}$, and
the excited state, $\ket{2}$, of the Hamiltonian (\ref{HamTLS}) have
energies  $E_{1,2} = \mp \frac{\Omega_0}{2}$, with the energy
splitting $\Omega_0 = \sqrt{\Delta^2 + h^2}.$ We assume that the
system-bath interaction is determined by the Hamiltonian
\ba H_{\rm int} = - Q\, \sigma_z, \ea
where $Q$ is a quantum-mechanical operator of the bath. The bath
itself has a Hamiltonian $H_B$, so that the total Hamiltonian $H$ of
the problem is a sum of three terms:
\ba \label{H1} H = H_S + H_{\rm int} + H_B. \ea
We assume that the free bath (with no coupling to the system) has a
Gaussian statistics \cite{ES81} determined by a spectrum of
fluctuations: \be
 S(\omega) = \int dt e^{i\omega t}\langle Q(t) Q(0) \rangle ,
\ee where $Q(t) = e^{iH_Bt} Q e^{-iH_Bt}$. Frequently, the Gaussian
bath is represented as a collection of harmonic oscillators
\cite{Leggett87}. Within our general formalism we do not have to
resort to any specific representation of the bath.

In many realistic situations (see, for example,
Refs.~\cite{Lanting11}), the noise may come from different sources,
some dominating at low frequencies (such as 1/f noise) and others
dominating at high frequencies. As such, we consider $S(\omega)$ to
be a sum of two terms:
\ba \label{SpLH} S(\omega)~=~S_L(\omega)~+~S_H(\omega),  \ea
where $S_L(\omega)$ and $S_H(\omega)$ are functions that are peaked
at low and high frequencies, respectively. Each function may tail
into the other function's region. Hereafter we refer to the noise
with the spectrum (\ref{SpLH}) as  the \emph{hybrid} noise.  The
formula of high-frequency spectrum $S_H$ is given in Sec. III E. For
the explicit expression for $S_L(\omega)$ we refer to Eq.~(B3) shown
in the Supplementary Information Section of Ref.~\cite{Johnson11},
although there are no need for these formulas here.  Notice also
that in the present paper we operate with the
experimentally-measured parameters of the low-frequency bath, such
as the noise intensity $W^2$ and the reorganization energy
$\varepsilon_L$, which are are defined below.

The relaxation dynamics of the qubit become simple in two
situations. First, when the qubit is weakly coupled to only a
high-frequency (HF) bath and the energy splitting of the qubit is
larger than the broadening of qubit's energy levels. In this case,
the relaxation is described by the Bloch-Redfield rate
\cite{Leggett87,Slichter90,Walls94},
\ba \label{GaBR} \Gamma = \frac{\Delta^2}{\Delta^2 + h^2} \,
S_H(\Omega_0),  \ea
which is valid when $\Gamma \ll \Omega_0$. Herein, we make the
following assumptions for the Boltzmann and Planck constants: $k_B =
1, \; \hbar = 1.$

The second case is when the qubit is coupled only  to a
low-frequency (LF) bath and its tunneling amplitude is much smaller
than the energy broadening caused by noise. The qubit dynamics
therefore becomes incoherent and the resulting macroscopic resonant
tunneling (MRT) rate is given by \cite{Amin08, AminBrito09}
\ba \label{GaM} \Gamma = \frac{\Delta^2}{8}\, \sqrt{\frac{2
\pi}{W^2}}\, \exp\left[ - \frac{(h - 4\, \varepsilon_L )^2}{ 8 W^2}
\right],  \ea
where
 \ba \label{eW} W^2 = \int \frac{d\omega}{2\pi} S_L(\omega),
\;\;\varepsilon_L = \int \frac{d\omega}{2\pi}
\frac{S_L(\omega)}{\omega}  \ea
determine the intensity of the noise and the reorganization energy
(the shift  of the bath energy due to the change of the qubit
state), respectively. The fluctuation-dissipation theorem leads to
$W^2 = 2 \,\varepsilon_L\, T,$ where $T$ is the equilibrium
temperature of the bath \cite{Amin08}. Equation (\ref{GaM}),
commonly known as  the Marcus formula
\cite{Marcus93,Cherepanov01,Yang02, Ghosh11}, is valid when the
tunneling amplitude $\Delta$ is much smaller than the MRT line-width
$W$: $\Delta \ll W.$  This equation has been successful in
explaining experimental data from flux qubits
\cite{Lanting11,Harris08}.

In practice, low and high-frequency noises coexist and both have to
be considered in the dynamics of the qubit. In
Refs.~\cite{Lanting11,Amin08} the formula (\ref{GaM}) has been
generalized to include effects of high-frequency noise on the MRT
rate. For small tunneling amplitudes $\Delta$  the modified rate
$\Gamma$ is described by the following integral:
\ba \label{GaLan}\Gamma = \frac{\Delta^2}{4} \, \int d\tau \; e^{i
(h - 4 \varepsilon_L) \tau - 2 W^2 \tau^2 } \times \nn  \exp\left[ 4
\int \frac{d\omega}{2 \pi} \,\frac{S_H(\omega)}{\omega^2} \, \left(
e^{- i \omega \tau} - 1 \right) \right].   \ea
We notice that the integrand of Eq.~(\ref{GaLan}) is equal to the
product of the low-frequency component, $ e^{-2 W^2 \tau^2 -4 i
\varepsilon_L \tau}$, multiplied by the high-frequency factor, which
depends on the spectrum $S_H$. The low-frequency component has the
Gaussian Fourier image,
\ba G^L(\omega) = \sqrt{\frac{\pi}{ 2 W^2} }\, \exp\left[ -
\frac{(\omega - 4 \varepsilon_L)^2 }{8 W^2} \right]. \ea
The high-frequency factor is characterized by the more complicated
integral:
\ba \label{GaH} G^H(\omega) = \int_{-\infty}^{+\infty} d \tau\;
e^{i\omega \tau} \times   \nn \exp\left[ 4 \int \frac{d\Omega}{2
\pi} \,\frac{S_H(\Omega)}{\Omega^2} \, \left( e^{- i \Omega \tau} -
1 \right) \,\right]. \ea
We notice that both functions, $G^L(\omega)$ and $G^H(\omega)$,
satisfy the normalization condition,
\ba \label{GaNorm} \int \frac{d\omega}{2\pi}\, G^\mu(\omega) = 1,
\ea
where $\mu = L, H.$

The rate (\ref{GaLan}) can be represented as a convolution of the
Gaussian envelope $G^L(\omega)$ and the function $G^H(\omega)$,
\ba \label{GCon} \Gamma = \frac{\Delta^2}{4}\, \int \frac{d
\omega}{2\pi}\, G^L(h - \omega)\, G^H(\omega). \ea
In the Markovian case, where the spectrum $S_H(\omega)$ is flat,
$S_H(\omega) = S_H(0)$, the function $G^H(\omega)$ has a Lorentzian
shape,
\ba \label{GLoR} G^H(\omega) = \frac{ 4 S_H(0)}{\omega^2 + [ 2
S_H(0) ]^2 }. \ea
Here we need not to assume that the qubit-bath coupling is small.
Equation~(\ref{GLoR}) is valid at frequencies $0 \leq \omega \leq
1/\tau_H, $ where $\tau_H$ is a correlation time of the
high-frequency fluctuations  described by the function
$S_H(\omega).$ Later we introduce a spectral density $S_H$ of the
Ohmic noise characterized by the correlation time $\tau_H \sim 1/T.$

The Bloch-Redfield limit is described by Eq.~(\ref{GaH}) with the
frequency $\omega$, which is much larger than the coupling to the
environment given by the spectrum $S_H(\omega)$:
$\frac{S_H(\omega)}{\omega } \ll 1. $ With this small parameter, we
can expand  the dissipative factor in Eq.~(\ref{GaH}). Now the
function $G^H(\omega)$ turns into the form
\ba \label{GBRed} G^H(\omega) = 4 \frac{ S_H(\omega)}{\omega^2}. \ea
Equations~(\ref{GLoR}) and (\ref{GBRed}) can be approximately
combined into one Lorentzian formula that has a frequency-dependent
numerator,
\ba \label{GaLR} G^H(\omega) = \frac{ 4 S_H(\omega)}{\omega^2 + [ 2
S_H(0) ]^2 }. \ea
The Markovian and Bloch-Redfield expressions follow from this
formula in the corresponding limits. Notice also that the function
(\ref{GaLR}) is normalized according to Eq.~(\ref{GaNorm}).

Thus, the single-qubit relaxation rate (\ref{GaLan})  can be
conveniently represented as a convolution of the Gaussian and
Lorentzian line shapes,
 \ba \label{GaMR}
 \Gamma= \frac{\Delta^2}{4}\,  \int \frac{d\omega}{2\pi} \, \frac{S_H(\omega)}{
 \omega^2 + [ 2 S_H(0) ]^2 } \times \nn
 \sqrt{\frac{\pi}{ 2 W^2} }\, \exp\left[ -
\frac{(h - \omega - 4 \varepsilon_L)^2 }{8 W^2} \right]. \ea
The convolution integral in (\ref{GaMR}) has a simple
interpretation. One can think of the low-frequency noise as a random
shift in energy bias: $h\to h+h_{\rm noise}$, where  $h_{\rm noise}$
has Gaussian distribution with variance of $2W$. The high-frequency
relaxation rate, given by the Lorentzian line-shape, will therefore
be shifted by $h_{\rm noise}$. Ensemble averaging over low-frequency
fluctuations will lead to a convolution integral similar to
(\ref{GCon}) and (\ref{GaMR}). The reorganization energy
$\varepsilon_L$ is a result of the action of the qubit on the
environment.

The intuitive description above holds beyond the validity of
Eq.~(\ref{GaMR}). In the next sections, we will generalize this
approach to multiqubit systems without resorting to the small
tunneling amplitude approximation.

\section{Definitions and notations}
\label{HamSec}

\subsection{ The Hamiltonian}

We are interested in dissipative evolution of a quantum annealer
\cite{Johnson11, Lanting14, Dickson13} treated as a system of $N$
qubits coupled to a heat bath. The qubits are described by the
Hamiltonian:
\ba \label{HS} H_S = {\cal A}(s) H_D + {\cal B}(s) H_P,  \ea
where $H_D$ and $H_P$ are the driving (tunneling) and problem
Hamiltonians defined as
\ba \label{HDP} H_D &=& - \frac{1}{2} \, \sum_\alpha \,
\Delta_{\alpha} \, \sigma_x^{\alpha} , \nn H_P &=&  \frac{1}{2}
\,\sum_{\alpha} h_{\alpha} \sigma_z^{\alpha} + \frac{1}{2}
\,\sum_{\alpha \neq \beta} J_{\alpha \beta} \,\sigma_z^{\alpha}
\sigma_z^{\beta}.  \ea
The energy functions ${\cal A}(s)$ and ${\cal B}(s)$ determine the
annealing schedule with $s = t/t_f$ being the dimensionless
annealing parameter ($ 0\leq s \leq 1$), and $t $ is and $t_f$ being
the running time and the total annealing time, respectively. Details
of the annealing schedule are unimportant for the current discussion
as long as the time-dependent Hamiltonian changes slowly, which is
exactly the case for quantum annealing algorithms.

We assume an interaction with a bath of the form:
\ba \label{Hint} H_{\rm int} = - \sum_{\alpha=1}^N Q_{\alpha} \,
\sigma_z^{\alpha}, \ea
with operators $Q_{\alpha}$ characterized by Gaussian statistics
with zero average values, $\langle Q_{\alpha}\rangle = 0.$ We also
suppose that different qubits, labeled as $\alpha$ and $\beta$,  are
coupled to statistically independent environments, such that
$\langle Q_{\alpha} Q_{\beta} \rangle = 0$ if $\alpha \neq \beta.$
This has been experimentally confirmed for flux qubits
\cite{Lanting10}.

\subsection{Schr\"odinger picture}

The total system-bath Hamiltonian $H$ written in the Schr\"odinger
representation has the form given by Eq.~(\ref{H1}). Here, the time
evolution of the system-bath can be described by the density matrix
$\rho_{SB} = \ket{\psi_{SB}}\bra{\psi_{SB}}$, where
$\ket{\psi_{SB}}$ is the system-bath wave function. The
time-evolution of $\rho_{SB}$ is governed by the von Neumann
equation,
\ba\label{rhoSB}  i  \dot \rho_{SB} = [ H_S + H_B + H_{\rm int},
\rho_{SB} ],  \ea
where $[A,B]$ means a commutator of operators $A$ and $B$.
 We assume that the initial system-bath matrix can be
factorized into the product
\ba \label{rhoF}\rho_{SB}(0) = \rho_S(0)\otimes \rho_B  \ea
of the initial density matrix of the qubits, $\rho_S(0)$, and the
equilibrium matrix of the bath \cite{Blum12},
\ba \label{rhoB} \rho_B = \frac{e^{-H_B/T}}{{\rm Tr}_B
(e^{-H_B/T})}. \ea
Here, ${\rm Tr}_B$ denotes a trace over bath variables, and $T$ is
the bath temperature.

With the unitary matrix $U_B = e^{- i\, H_B t}$, the Hamiltonian $H$
(\ref{H1}) turns into the form
\ba \label{H2} H' &=& U_B^\dag\,( H_S + H_{\rm int} + H_B ) U_B -
i\, U_B^\dag \frac{\partial}{\partial t}\, U_B \nn &=& H_S
-\sum_\alpha Q_\alpha(t)\, \sigma^z_\alpha,  \ea
where $ Q_\alpha(t) $ is the free-evolving bath operator,
\ba \label{Qa} Q_\alpha(t) = e^{i H_B t}\,  Q_\alpha \, e^{-i H_B
t}.  \ea
The evolution of the density matrix can now be defined in terms of
the unitary operator \ba \label{UIa} U(t) = {\cal T}\, e^{-
i\,\int_0^t d\tau\, H'(\tau) }.  \ea
Hereafter, for simplicity of notation, we remove time dependences
from unitary matrices. A consecutive application of the operators
$U_B$ and $U$ produces the system-bath density matrix $\rho_{SB}(t)$
at time $t$,
\ba \label{rhoSBa} \rho_{SB}(t) = U_B U \rho_{SB}(0) U^\dag
U_B^\dag.  \ea
This time-dependent matrix presents the solution of the von Neumann
equation (\ref{rhoSB}). The average value of an arbitrary
Schr\"odinger operator ${\cal O}$, which describes a physical
variable of the qubits or of the bath, is determined by the
 density matrix $\rho_{SB}(t)$ (\ref{rhoSBa}) taken at
 time $t$,
\ba \label{sAv} \langle {\cal O} \rangle_{SB}(t) = {\rm Tr} [
\,\rho_{SB}(t) \, {\cal O}\, ] = \nn {\rm Tr}_B \sum_k \langle k\, |
\,\rho_{SB}(0) \, U^\dag\, U_B^\dag \,{\cal O} \,U_B\, U \,|\, k
\rangle. \ea
Here the total trace ${\rm Tr}$ includes the trace ${\rm Tr}_B$ over
free-bath variables and also the trace ${\rm Tr}_S$,
\ba \label{TrS} {\rm Tr}_S = \sum_k \langle k | \ldots | k \rangle,
  \ea
over a full set of qubit states $\{\ket{k}\}.$

\subsection{Heisenberg picture}

In the density matrix approach, the state of the system is described
via the reduced density matrix, which is obtained by averaging
$\rho_{SB}$ over the bath fluctuations. Some information about
quantum fluctuations is lost after the averaging. This limits the
method to calculations of only the averages and same-time
correlation functions. Other properties such as different-time
correlations remain beyond the reach of this approach. In the
Heisenberg picture, the equations are written in terms of the
operators without taking averages. This allows calculations of
correlation functions to any order as long as the equations can be
solved.

In the Heisenberg representation, the average value of an arbitrary
operator ${\cal O}$ in (\ref{sAv}) can be written as
\ba \label{sAvH} \langle {\cal O} \rangle_{SB}(t) = {\rm Tr} [
\,\rho_{SB}(0) \,{\cal O}^H(t) \,], \ea
where
\ba \label{sH} {\cal O}^H(t) = U^\dag U_B^\dag \,{\cal O} \,U_B U
 \ea
is the Heisenberg operator of the variable ${\cal O}.$ The
Schr\"odinger operator ${\cal O} $ may explicitly depends on time.
In this case, its partial derivative over time, $\frac{\partial
{\cal O}}{\partial t}$, is not equal to zero. It follows from
Eq.~(\ref{sH}) that the time evolution of the operator ${\cal
O}^H(t)$ is described by the Heisenberg equation
\ba \label{dsdt} i \, \frac{d}{dt}\, {\cal O}^H = [\, {\cal O}^H,
H^H \,] + (U_B U)^\dag i \;\frac{\partial {\cal O}}{\partial t}
\;U_B U, \ea
where the total Hamiltonian (\ref{H1}) is written in the Heisenberg
picture as
\ba \label{Ha}
 H^H = U^\dag U_B^\dag \,H \,U_B U   \ea

\subsection{The bath}

\label{SecBath} We assume that the bath coupled to $\alpha-$qubit is
described by Gaussian statistics \cite{ES81}. These statistics are
characterized by a correlation function $K_{\alpha}(t,t')$
\ba \label{KS} K_{\alpha}(t,t') = \langle Q_{\alpha}(t)
Q_{\alpha}(t') \rangle .  \ea
Here $Q_{\alpha}(t)$ is a free-evolving bath operator (\ref{Qa}).
The brackets $ \left< \ldots \right>$ denote the average of ${\cal
O}$  over the free-bath fluctuations,
\ba \label{TrB} \left< {\cal O} \right> = {\rm Tr}_B [\, \rho_B \,
{\cal O} \,],  \ea
unless otherwise specified. For stationary processes,
$K_{\alpha}(t,t')$ depends on the time difference, hence allowing
the spectral density to be defined as
\ba S_\alpha(\omega) = \int dt  e^{i \omega t} K_{\alpha}(t). \ea
In addition to the correlator (\ref{KS}), we introduce dissipative
functions $f_{\alpha}(t)$ and $g_{\alpha}(t)$ defined as
\ba \label{fgA} f_{\alpha}(t) = \int \frac{d \omega}{2 \pi} \,
\frac{S_{\alpha}(\omega)}{\omega^2} \, (1 - e^{- i \omega t }), \nn
g_{\alpha}(t) =  - i\, \dot f_{\alpha}(t) =  \int \frac{d \omega}{2
\pi} \, \frac{S_{\alpha}(\omega)}{\omega} \, e^{- i \omega t }.
 \ea
Notice that $K_{\alpha}(t)~=~\ddot f_{\alpha}(t).$ The total
reorganization energy of the bath is defined as \ba \label{epA}
\varepsilon_\alpha = \int \frac{d \omega}{2 \pi} \,
\frac{S_{\alpha}(\omega)}{\omega}.  \;\ea

The response of the bath to an external field is described by the
retarded Green function
\ba \label{phi1} \varphi_{\alpha}(t{-}t') = \langle i [
Q_{\alpha}(t), Q_{\alpha}(t') ] \rangle \, \theta(t{-}t').
 \ea
The causality is provided by the Heaviside step function
$\theta(t-t').$ The response function $\varphi_{\alpha}$ is related
to the susceptibility of the bath defined through
\ba \label{phi2} \chi_{\alpha}(\omega) = \int d\tau \, e^{i \omega
\tau }\varphi_{\alpha}(\tau).  \ea
According to the fluctuation-dissipation theorem, in equilibrium
$S_{\alpha}(\omega)$ is proportional to the imaginary part
$\chi_{\alpha}''(\omega)$ of the bath susceptibility,
\ba \label{FDT} S_{\alpha}(\omega) = \chi_{\alpha}''(\omega)\,
\left[ \coth\left(\frac{\omega}{2 T}\right) + 1 \right],   \ea
where $T$ is the temperature of the equilibrium bath.

\subsection{Hybrid noise}
\label{HybridNoise}

 Hereafter we assume that the dissipative
environments coupled to different qubits, although uncorrelated,
have the same spectral density of bath fluctuations:
$S_\alpha(\omega) = S(\omega)$. The same is true of the functions
$f_\alpha(\tau) = f(\tau)$, $g_\alpha(\tau) = g(\tau)$, and
$\chi_\alpha(\omega) = \chi(\omega).$ In the case of hybrid noise,
$S(\omega)$  is given by Eq.~(\ref{SpLH}).
 The dissipative functions $f$
and $g$  can be split into low and high-frequency  components,
\ba \label{fgB} f = f_L + f_H, \qquad g = g_L + g_H.  \ea
For the low-frequency part of the function $f$, one can expand
$e^{i\omega \tau}$ in  Eq.~(\ref{fgA}), assuming $\omega\tau \ll 1$.
Keeping up the second order in $\omega\tau$, we obtain
\ba \label{fL} f_L(\tau ) = i\, \varepsilon_L \, \tau +
\frac{1}{2}\, W^2 \, \tau^2,  \ea
with $\varepsilon_L$ and $W$ defined in Eq.~(\ref{eW}). 

To treat the high-frequency parts, we assume Ohmic noise
\ba \label{SH} S_H(\omega) = \frac{\eta \omega}{ 1 - e^{-\omega/T}}
\, e^{-|\omega|/\omega_c}.  \ea
Here $\eta$ is a small dimensionless coupling constant  and
$\omega_c$ is a large cutting frequency of the high-frequency noise.
This assumption is justified experimentally  \cite{Lanting11} and
also theoretically  \cite{Leggett87}. The dissipative functions
$f_H$ and $g_H$ are calculated in Appendix~\ref{ApxDF}. The total
reorganization energy, $\varepsilon_\alpha \equiv \varepsilon,$ is
defined by (\ref{epA}), so that $\varepsilon = \varepsilon_L +
\varepsilon_H.$ Here $\varepsilon_L$ is defined in (\ref{eW}), and
$\varepsilon_H$ is the high-frequency component of the
reorganization energy (\ref{epA}). For the Ohmic spectrum (\ref{SH})
of the bath, we have $\varepsilon_H = \frac{\eta \omega_c}{2 \pi}.$

\subsection{Selection of the basis}
\label{Basis}

The dynamical equations we aim to derive must be represented in a
convenient basis, which we denote by $\{\ket{n(t)}\}$. This basis
could be the instantaneous eigenstates of the system Hamiltonian or
some superpositions of those. The system-bath Hamiltonian $H'$ in
(\ref{H2}) can be written as
\ba \label{H4} &&H' = \sum_n [ \,E_n - Q_n(t) \,]\, \ket{n}\bra{n} +
\nn &&\sum_{m\neq n} [ \,T_{mn} - Q_{mn}(t) \,]\,\ket{m}\bra{n}.
 \ea
where
\ba \label{ETmn} E_n = \langle n | H_S | n \rangle, \quad T_{mn} =
\langle m | H_S | n \rangle,  \ea
\ba \label{Qmn} &&Q_n(t) = \sum_{\alpha=1}^N \sigma^\alpha_n\,
Q_\alpha(t), \nn &&Q_{mn}(t) = \sum_{\alpha=1}^N
\sigma^\alpha_{mn}\, Q_\alpha(t),  \ea
with $Q_\alpha(t)$ defined in  (\ref{Qa}), and
\ba \label{sAmn} \sigma^\alpha_n = \langle n | \sigma_z^\alpha | n
\rangle, \;\; \sigma^\alpha_{mn} = \langle m | \sigma_z^\alpha | n
\rangle.  \ea
We also introduce the following notations, which we will use later:
\ba \label{abc} a_{mn} = \sum_{\alpha} (\sigma_m^\alpha -
\sigma_n^\alpha)^2, \; b_{mn} = \sum_\alpha |\sigma^\alpha_{mn}|^2,
\hspace{0.75cm}
 \\ c_{mn} = \sum_\alpha \sigma^\alpha_{mn} (\sigma_m^\alpha
- \sigma_n^\alpha), \; d_{mn} = \sum_\alpha \sigma^\alpha_{mn}
(\sigma_m^\alpha +\sigma_n^\alpha). \nonumber \ea
Hereafter, we will refer to the parameter $a_{mn}$ as to the Hamming
distance between states $\ket{m}$ and $\ket{n}$. We also notice that
the parameters $a_{mn}$ and $b_{mn}$ are real and positive, and
$c^*_{mn} = - \,c_{nm}, \; d_{mn}^* = d_{nm}. $

\section{System evolution in the interaction representation}
\label{IntPic}

Properties of the system of qubits are determined by the reduced
density matrix
\ba \label{rhoSa} \rho_S =  {\rm Tr}_B \rho_{SB} = {\rm Tr}_B [ U
\rho_{SB}(0) U^\dag ], \ea
where the system-bath density matrix $\rho_{SB}$ is given by
Eq.~(\ref{rhoSBa}). A system-bath average of an arbitrary  operator
${\cal O}_S$ of the system is written as
\ba \langle {\cal O} \rangle _{SB} = {\rm Tr}_S [ \rho_S(t) {\cal
O}_S ]. \ea
In the basis  introduced in Sec.~\ref{Basis}, the matrix $\rho_S$
has the form
\ba \label{rhoSn} \rho_S = \sum_{mn} \rho_{nm} \ket{n}\bra{m}, \ea
with the matrix elements defined as
\ba \label{rhoMN} \rho_{nm} = \langle n | \rho_S | m \rangle. \ea
Our goal is to derive a set of master equations for the probability
distribution  of the qubits, $P_n$, over the states $\{\ket{n}\}$,
where
\ba \label{Pna} P_n = \rho_{nn} = \langle n | \rho_S | n \rangle.
\ea
The time evolution of the matrix (\ref{rhoSa}) is determined by the
unitary operator $U$ defined by (\ref{UIa}) where the Hamiltonian
$H'$ is given by Eq.~(\ref{H4}).  The objective is to go beyond the
perturbation theory in the system-bath coupling. This can be done by
treating $Q_n$ exactly, but $Q_{mn}$ perturbatively. The interaction
representation is best suited for this goal.

A transition to the interaction picture, although straightforward
for time-independent bases, becomes more involved if the basis
changes in time. Let us introduce a unitary operator
\ba \label{Ua} U_0(t) = \sum_n e^{- i  \phi_n(t) } \, {\cal S}_n(t)
\, \ket{n}\bra{n}, \ea
where $\ket{n}$ is a time-dependent basis of the system, and
\ba \label{phin}
 \phi_n(t) = \int_0^t d\tau
E_n(\tau)  \ea
is written in terms of average energies $E_n(t)~=~\langle n(t) |
H_S(t) | n(t) \rangle$. We also introduce the $S$-matrix:
\ba \label{snA} {\cal S}_n(t) = {\cal T} \exp\left[ i \sum_\alpha
\sigma_n^\alpha(t) \int_0^t dt_1\, Q_\alpha(t_1) \right], \ea
with ${\cal T}$ being the time-ordering operator for $t_1$. Notice
that the time-dependent matrix element $\sigma_n^\alpha(t)$ is taken
out of the integral over $t_1$. This becomes necessary when we want
to express correlation functions in terms of dissipative functions.

The interaction Hamiltonian is given by the expression
\ba \label{HIa} H_I = U_0^\dag H' U_0 - i U_0^\dag \dot U_0. \ea
This Hamiltonian defines  the unitary evolution operator
\ba \label{UI} U_I(t) = {\cal T} e^{- i \int_0^t d\tau H_I(\tau) }.
 \ea
We expect that the Hamiltonian $H_I$ does not contain the
nonperturbative diagonal terms $Q_n$. To calculate the
time-derivative $\dot U_0$ in (\ref{HIa}), we need
\ba \label{sndot} - i \dot{\cal S}_n(t) = Q_n(t) \, {\cal S}_n(t) +
 \nn \sum_\alpha \dot \sigma_n^\alpha (t) \, {\cal S}_n(t)
\, \int_0^t d\tau \, \tilde {\cal S}_n^\dag(t,\tau) Q_\alpha(\tau)
\, \tilde {\cal S}_n(t,\tau), \ea
where
\ba \label{snB} \tilde  {\cal S}_n(t,\tau) = {\cal T} \exp\left[ i
\sum_\alpha \sigma_n^\alpha(t) \int_0^\tau dt_1\, Q_\alpha(t_1)
\right]. \ea
Notice that $ {\cal S}_n(t)= \tilde {\cal S}_n(t,t)$. In the
interaction picture, the system-bath Hamiltonian $H_I$ (\ref{HIa})
 takes the form
\ba \label{HIc} H_I = i \, \sum_{n}  | \dot n\rangle \bra{n} -
\sum_{mn} \tilde Q_{mn}(t)\, \ket{m(t)}\bra{n(t)}.
 \ea
The modified bath operator $\tilde Q_{mn}$ has diagonal terms
\ba \label{QTn} \tilde Q_{nn}(t) =   \nn - \sum_\alpha \dot
\sigma_n^\alpha (t) \int_0^t d\tau \, \tilde{\cal S}_n^\dag(t,\tau)
\, Q_\alpha(\tau) \, \tilde{\cal S}_n(t,\tau) , \hspace{0.25cm} \ea
and off-diagonal ($m\neq n$) terms,
\ba \label{QTmn} \tilde Q_{mn} &=& e^{i \phi_{mn}(t)} \,{\cal
S}_m^\dag(t)\, [  Q_{mn}(t) - \tilde T_{mn} ] \, {\cal S}_n(t).
 \ea
Here
\ba \label{phiMN} \phi_{mn}(t) = \phi_m(t) - \phi_n(t) = \int_0^t d
\tau \, \omega_{mn}(\tau), \ea
is defined in terms of
\ba \label{omA} \omega_{mn}(t) = E_m(t) - E_n(t), \ea
and
\ba \label{dotSz} \dot \sigma_n^\alpha (t) = \sum_{m\neq n} [\;
\sigma_{mn}^\alpha \langle \dot n | m\rangle + \langle m | \dot n
\rangle\, \sigma^\alpha_{nm} \;]. \ea
We also introduce
\ba \label{TmnT} \tilde T_{mn} = T_{mn} - i \langle m | \dot n
\rangle , \ea
with $T_{mn}$ defined in (\ref{ETmn}). When the basis $\{\ket{n}\}$
is formed by the instantaneous eigenstates of the Hamiltonian $H_S$,
$ H_S \ket{n} = E_n \ket{n},$ we obtain
\ba \label{MdotN} \langle m| \dot n\rangle = \frac{1}{t_f} \, \frac{
\langle m (s) | \frac{d H_S(s)}{ds} | n(s)\rangle }{E_n(s) -
E_m(s)}. \ea
 Here we assume that the spectrum
$E_n$ is nondegenerate and that $H_S$ (\ref{HS}) is characterized by
real parameters. In this case we have $\langle n| \dot n\rangle =
0.$

In Appendix \ref{ApxQC}, we calculate correlation functions $\tilde
K_{mn}^{m'n'}(t,t')$ of the bath variables (\ref{QTmn}),
\ba \label{KmnA} \tilde K_{mn}^{m'n'}(t,t') = \langle \tilde
Q_{mn}(t), \tilde Q_{m'n'}(t') \rangle. \ea
We show that the only terms that survive during the annealing run
are characterized by the relation
\ba \label{KmnB} \tilde K_{mn}^{m'n'}(t,t') = \delta_{m n'}
\delta_{n m'} \, \tilde K_{mn}(t,t'), \ea
where the function $\tilde K_{mn}(t,t')$ is given by
Eq.~(\ref{Cor7a}). In addition, we demonstrate that, during
annealing, correlations between diagonal operators $\tilde Q_{kk}$
(\ref{QTn}) and off-diagonal  bath variables $\tilde Q_{mn}$
(\ref{QTmn}) rapidly disappear in time, such that  $\langle \tilde
Q_{mn}(t), \tilde Q_{kk}(t') \rangle \sim 0.$ The same is true for
the average values of the operators (\ref{QTmn}): $\langle \tilde
Q_{mn}(t)\rangle  \sim 0.$

\subsection{Time evolution}

The evolution of the matrix $\rho_S$ is determined by the unitary
matrix (\ref{UIa}), which can be written as: $U = U_0 U_I$, where
$U_0$ and $U_I$ are given by Eqs.~(\ref{Ua}) and (\ref{UI}). In the
interaction picture, we have
\ba \label{rhoNM} \rho_{nm} &=& e^{ i \phi_{mn}} {\rm Tr} [
\rho_{SB}(0) U_I^\dag \ket{m}{\cal S}_m^\dag {\cal S}_n \bra{n} U_I
] \nn &=& e^{ i \phi_{mn}} {\rm Tr} [ \rho_{SB}(0) U_I^\dag {\cal
S}_m^\dag {\cal S}_n U_I \Lambda_{mn} ], \nn &=& e^{ i \phi_{mn}}
{\rm Tr}_S [ \rho_{S}(0) \langle U_I^\dag {\cal S}_m^\dag {\cal S}_n
U_I \Lambda_{mn} \rangle ]. \ea
Here, we have used  (\ref{rhoF},\ref{rhoSa},\ref{rhoMN}) and have
introduced the interaction picture operator
\ba \label{rhoTa} \Lambda_{mn} = U_I^\dag \, \ket{m}\bra{n}\, U_I,
 \ea
which will play an important role in our theory. The bath average
$\langle ... \rangle$ is defined in  (\ref{TrB}). Equation
(\ref{rhoNM}) becomes simplified for the diagonal elements:
\ba \label{Pnb} P_n = \rho_{nn}  =  {\rm Tr}_S [ \rho_S(0) \langle
\Lambda_{nn} \rangle ]. \ea
In the following, we consider time evolution of the operators
$\Lambda_{nn}$ instead of working with the elements (\ref{rhoNM})
and (\ref{Pnb}) of the system density matrix. Working with operators
instead of averages allows derivation of more accurate master
equations.

Taking the derivative of (\ref{rhoTa}) and using (\ref{HIc}), we
obtain
\ba \label{rTa} i \frac{d}{dt} \,\Lambda_{mn} =\sum_{k} ( Q_{km}^I
\, \Lambda_{kn} - Q_{nk}^I \, \Lambda_{mk}),  \ea where
\ba \label{QTa}  Q_{mn}^I = U_I^\dag\, \tilde Q_{mn} \, U_I. \ea
Here we use the fact that $\Lambda_{mn}$ and $ Q_{kl}^I$, taken at
the same moment of time $t$, commute: $[ \Lambda_{mn}, Q_{kl}^I
]=0,$ for any set of indexes $m,n,k,l.$ The evolution of the
diagonal elements $\Lambda_{nn}$ is of prime interest since these
elements determine the probabilities (\ref{Pnb}):
\ba \label{rTb} i \frac{d}{dt} \, \Lambda_{nn} =  \sum_{m\neq n} (
 Q_{mn}^I \, \Lambda_{mn} -   Q_{nm}^I \, \Lambda_{nm}
).  \ea
Notice that the diagonal elements of the bath, $Q_{mm}^I $ and
$Q_{nn}^I$, have no influence on the evolution of $\Lambda_{nn}$.
Averaging over free bath fluctuations leads to
\ba \label{rT1} i \frac{d}{dt} \, \langle \Lambda_{nn}\rangle  =
\sum_{m\neq n} ( \langle Q^I_{mn} \, \Lambda_{mn}\rangle - \langle
Q^I_{nm} \, \Lambda_{nm}\rangle  ). \ea
This equation is exact and difficult to solve without
approximations.

To simplify Eq.~(\ref{rT1}), we use perturbation expansion assuming
that $\tilde Q_{mn}$ is small. Appendix \ref{ApxMeq} shows that the
probability distribution $P_n$ of the system (\ref{Pnb}) follows the
master equation:
\ba \label{rT5a}   \dot P_n  + \Gamma_n P_n = \sum_m \Gamma_{nm} \,
P_m , \ea
where $\Gamma_n = \sum_m\Gamma_{mn}$ and
\ba \label{GamB} \Gamma_{nm} = \int_{-\infty}^{+\infty} d\tau\, e^{i
\omega_{mn} \tau - a_{mn} f(\tau)} \times \nn \{ b_{mn} \ddot
f(\tau) + [ \bar T_{mn} - c_{mn} g(\tau) ] \,[ \bar T_{mn}^* -
c_{mn}^* g(\tau) ] \}.
 \ea
Coefficients  $a_{mn}, b_{mn}, d_{mn}$ are defined by
Eq.~(\ref{abc}) and
\ba \label{TkC} \bar T_{mn}  = T_{mn} - i \langle m | \dot n \rangle
- d_{mn} \,\varepsilon,    \ea
where $\varepsilon$ is the total reorganization energy described in
Sec.~\ref{HybridNoise}, $\bar{T}_{mn}^* = \bar{T}_{nm}.$ All matrix
elements of the system operators in Eq.~(\ref{GamB}) are taken at
the running moment of time $t$.

 The rate $\Gamma_{nm}$ can be written in a form  similar to the single-qubit expression
 (\ref{GaLan}) and also to the multiqubit rate $\Gamma_{1 \rightarrow 0}$ given by Eq.~(5)
 from Ref. \cite{Boixo16} and by Eq.~(68) from Ref.~\cite{Boixo14},
\ba \label{GamTime} \Gamma_{nm} = \int_{-\infty}^{+\infty} d \tau \;
e^{i \omega_{mn} \tau}\,  e^{- i \varepsilon_{mn} \tau - \frac{1}{2}
W^2_{mn} \tau^2 } \times \nn \left[ ( 1 + i \omega_c \tau ) \frac{
\sinh(\pi T \tau)}{\pi T \tau} \right] ^{- \frac{\eta_{mn}}{2 \pi} }
\times \nn \{ b_{mn} \ddot f(\tau) + [ \bar T_{mn} - c_{mn} g(\tau)
] \,[ \bar T_{mn}^* - c_{mn}^* g(\tau) ] \}.
 \ea
Here we use Eqs.~(\ref{fL}) and (\ref{fH}) and introduce the
following parameters:
\ba \label{Wmn} \varepsilon_{mn} = a_{mn} \varepsilon_L, \, W^2_{mn}
= a_{mn} W^2,\, \eta_{mn} = a_{mn} \eta. \ea
We notice that, compared to previous results (see  Eqs.~(5), (6) in
\cite{Boixo16} and Eqs.~(43),(52),(54),(68) in \cite{Boixo14}), the
rate (\ref{GamTime}) does not contain any polaron shifts to the
frequency $\omega_{mn}$. Moreover, we have no need to represent the
bath as a system of harmonic oscillators as  done in Refs.
\cite{Boixo16} and \cite{Boixo14}.

\subsection{Applicability conditions}

The master equations (\ref{rT5a}) have been derived in
Appendix~\ref{ApxMeq} with the proviso that
\ba \label{GamTau} \Gamma_{nm}  \tau_{mn}  \ll 1, \ea
where $\Gamma_{nm}$ is the relaxation rate (\ref{GamB}). The inverse
correlation time of the bath, $\tau_{mn}^{-1}$, is estimated in
Appendix~\ref{ApxQC} as the maximum of two parameters: the average
energy distance  $|E_m - E_n|$ between the states $\ket{m}$ and
$\ket{n}$ and the MRT line-width $W_{mn} = W \sqrt{a_{mn}}$,
\ba \label{tauC} \frac{1}{\tau_{mn}} = {\rm max} \{ |E_m - E_n|,
W_{mn} \}.   \ea

\section{Relaxation rate as a convolution of Bloch-Redfield and
Marcus envelopes} \label{BRM}

In this section we show that, in addition to the expression
(\ref{GamTime}) of the rate $\Gamma_{nm}$ as the integral over time,
the same rate can be conveniently represented as a convolution
integral over frequencies of the Gaussian envelope multiplied by the
Lorentzian function. As in the case of a single qubit described in
Sec.~\ref{SingleQ}, the Gaussian curve is produced by low-frequency
bath noise. The Lorentzian factor is due to effects of the
high-frequency environment. Here, our aim is a generalization of the
single-qubit formula (\ref{GaMR}) to the multi-qubit case where
both, low-frequency noise and the single-qubit tunneling, can be
large.

\subsection{Convolution form of the rate $\Gamma_{nm}$} \label{ConGL}

Using integration by parts and $\dot f(\tau)=ig(\tau)$, we obtain
\ba \int d\tau\; e^{i \omega \tau}  e^{-a_{mn} f(\tau)}  g(\tau) =
\frac{\omega}{a_{mn}}  \int d\tau \; e^{i \omega \tau} e^{-a_{mn}
f(\tau)},\nonumber \ea
and
\ba \int d\tau \; e^{i \omega \tau}  e^{-a_{mn} f(\tau)} g^2(\tau) =
\nn \int d\tau \;e^{i \omega \tau} e^{-a_{mn} f(\tau)} \left[
\left(\frac{\omega}{a_{mn}}\right)^2 - \frac{1}{a_{mn}}\,\ddot
f(\tau) \right]. \nonumber \ea
Equation (\ref{GamB}) can therefore be represented  as
\ba \label{Gx2} \Gamma_{nm} =   \;\int d\tau \,e^{i \,\omega_{mn}
\tau} \, e^{-a_{mn} \,f(\tau)} \times \hspace{0.5cm} \nn \left[
 \left( b_{mn} -
\frac{|c_{mn}|^2}{a_{mn}} \right)\ddot f(\tau) + \left| \bar T_{mn}
- \omega_{mn} \, \frac{c_{mn}}{a_{mn}} \right|^2 \right].
\hspace{0.25cm} \ea

Let us introduce Fourier transformations
\ba \label{GLH} G^{\mu}_{mn}(\omega) = \int_{-\infty}^{\infty}\,
d\tau \, e^{i\omega \tau} \; e^{-a_{mn} f_{\mu}(\tau)}, \ea
where $\mu = L,H$ for  low and high-frequency noise, respectively.
Our goal is to write (\ref{Gx2}) as a convolution of the two
functions $ G_{mn}^L(\omega)$ and $ G_{mn}^H(\omega)$. The integrand
of Eq.~(\ref{Gx2}) contains a term $e^{- a_{mn} f(\tau)}$, which can
be written in the following form,
\ba e^{- a_{mn} f(\tau)} = e^{- a_{mn} f_L(\tau)} e^{- a_{mn}
f_H(\tau)} = \nn \int \frac{d \omega_1}{2\pi} \int \frac{d
\omega_2}{2\pi} \,e^{- i (\omega_1 + \omega_2) \tau}
\,G_{mn}^L(\omega_1)\, G_{mn}^H(\omega_2). \nonumber \ea
Substituting in (\ref{Gx2}) and taking the integral over $\tau$, we
obtain:
\ba \label{GamD} \Gamma_{nm} =  \int
\frac{d\omega}{2\pi}\;\Delta^2_{mn}(\omega)\,
G_{mn}^L(\omega_{mn}-\omega)\, G_{mn}^H(\omega), \ea
where
\ba \label{DeKN} \Delta^2_{mn}(\omega) =
  | A_{mn}|^2 +
B_{mn}\, (\omega^2 + W^2_{mn}),  \ea
with
\ba \label{AB} A_{mn} =   \bar T_{mn} - \omega_{mn} \,
\frac{c_{mn}}{a_{mn}},
\nn B_{mn} = \frac{a_{mn} b_{mn} - |c_{mn}|^2}{a_{mn}^2} =  \nn
\frac{1}{2 a_{mn}^2} \; \sum_{\alpha \beta} | (\sigma_m^{\alpha} -
\sigma_n^{\alpha} ) \,\sigma_{mn}^{\beta} - (\sigma_m^{\beta} -
\sigma_n^{\beta} ) \,\sigma_{mn}^{\alpha} |^2. \ea
Here, we have used $\ddot f = W^2 + \ddot f_H$ and neglected $\dot
f_H^2$, which is $O(\eta^2)$ in the weak coupling approximation.
Notice that $B_{mn}$ is always positive and disappears in a
single-qubit case where $\alpha = \beta = 1.$ Also, we have
$A_{mn}^* = A_{nm}$ and $ B_{nm} = B_{mn}.$

Appendix~\ref{ApxGL} shows that the low-frequency function
$G_{mn}^L(\omega)$ has a Gaussian shape,
\ba \label{GLa} G^L_{mn}(\omega) =  \sqrt{\frac{2 \pi}{W^2_{mn}} }\,
\exp \left[ - \frac{(\omega - \varepsilon_{mn} )^2}{ 2 W^2_{mn}}
\right].   \ea
A similar line shape describes the rate of macroscopic resonant
tunneling (MRT) in a system of qubits \cite{Amin08}. The
high-frequency component $G_{mn}^H(\omega)$ can be approximated by a
Lorentzian form, combining both Bloch-Redfield and Markovian rates:
\ba \label{GHa}
 G_{mn}^H(\omega) = \frac{a_{mn}\, S_H(\omega)}{\omega^2 +
\gamma_{mn}^2}. \ea
The parameter $\gamma_{mn} = \frac{a_{mn}}{2}\, S_H(0)$ does not
depend on frequency $\omega$. It follows from Eq.~(\ref{SH}) that
$S_H(0) = \eta T.$

\section{Special cases}
\label{Limits}

In this section we verify detailed balance conditions for the
equilibrium distribution of qubits and also consider the
Bloch-Redfield and Marcus limits of the rates (\ref{GamD}). In
addition, we apply the results of the previous section to a single
qubit interacting with a hybrid environment.

\subsection{Equilibrium condition}

We conclude from Eq.~(\ref{GamD}) that
\ba \label{GeqB} \Gamma_{mn} = \exp\left(
-\frac{\omega_{mn}}{T}\right) \, \Gamma_{nm},  \ea
where $\omega_{mn}$ is defined by Eqs.~(\ref{ETmn}) and (\ref{omA}).
 It follows from Eq.~(\ref{rT5a}) that the equilibrium
probabilities $P_n^{\rm eq}$ and $P_m^{\rm eq}$ to observe the
qubits in the states $\ket{n}$ and $\ket{m}$, respectively, obey the
equation:
\ba \label{PeqA} \sum_m (\Gamma_{mn} P_n^{\rm eq} - \Gamma_{nm}
P_m^{\rm eq} ) = 0.  \ea
The solution of this equation follows the detailed balance
condition:
\ba \label{PeqB} \frac{ P_m^{\rm eq} }{ P_n^{\rm eq} } = \exp\left[
- \frac{E_m - E_n }{T} \right],  \ea
with the local energy levels $E_m$ and $E_n$ (\ref{ETmn}) and the
bath temperature $T$.

The set of master equations (\ref{rT5a}) with the rates
$\Gamma_{nm}$ given by Eq.~(\ref{GamD}) provides a description of
the dissipative dynamics of a quantum annealer during the entire
annealing process. This description should be complemented by the
equation for the off-diagonal elements $\rho_{nm}$ of the system
density matrix. The time evolution of $\rho_{nm}$ is approximately
described by the formula
\ba \label{rNM} \rho_{nm} = e^{i \phi_{mn}} \langle {\cal S}_m^\dag
{\cal S}_n \rangle {\rm Tr}_S [ \rho_S(0) \langle \Lambda_{mn}
\rangle ] \simeq \nn e^{i \phi_{mn}(t)} \langle {\cal S}_m^\dag(t)
{\cal S}_n(t) \rangle {\rm Tr}_S [ \rho_S(0) \langle \Lambda_{mn}(0)
\rangle ]  . \ea
To derive this relation, we start with Eq.~ (\ref{rhoNM}) and move
out the dephasing factor $\langle {\cal S}_m^\dag(t) {\cal S}_n(t)
\rangle $ assuming that the matrices  ${\cal S}_m^\dag$ and $ {\cal
S}_n$ are weakly correlated with the operator $\Lambda_{mn} $
(\ref{rhoTa}). In Eq.~(\ref{rNM}) we have two possibilities: in the
first case the energy gap between states $\ket{m}$ and $\ket{n}$ is
large, therefore the factor $e^{i\phi_{mn}} \simeq e^{i \omega_{mn}
t}$ rapidly oscillates in time; in the second case the factor
$\langle {\cal S}_m^\dag(t) {\cal S}_n(t) \rangle $, which is given
by Eq.~(\ref{SmSn}), is the fast-decaying function of time. In both
cases, the correlation time $\tau_{mn}$ defined by Eq.~(\ref{tauC})
is much shorter than the time scale $\Gamma_{nm}^{-1}$  of the
variables $\langle \Lambda_{mn}\rangle $ and $\langle
\Lambda_{nn}\rangle $. Therefore, in Eq.~(\ref{rNM})  the function
$\langle \Lambda_{mn}(t) \rangle$  can be replaced by its initial
value $\langle \Lambda_{mn}(0) \rangle$.  Equation~(\ref{rNM})
describes fast dephasing of the system of qubits.

\subsection{Bloch-Redfield and Marcus limits}

For the qubits weakly interacting with the high-frequency noise in
the absence of low-frequency noise, the parameters of the
low-frequency bath go to zero: $W = 0, \varepsilon_L = 0.$ The
Gaussian envelope $G^L_{mn}(\omega)$ (\ref{GLa}) is approximated by
the function $ 2 \pi \delta(\omega)$, and the rate $\Gamma_{mn}$
(\ref{GamD}) takes the form
\ba \label{GamBR} \Gamma_{nm}^R = \Delta^2_{mn}(\omega_{mn})\,
\frac{ a_{mn}\, S_H(\omega_{mn})}{\omega_{mn}^2 + \gamma_{mn}^2.}.
\ea
 At a sufficiently large distance $\omega_{mn}$ between the
energy levels $E_m$ and $E_n$,  we find that
$$\Delta^2_{mn}(\omega_{mn}) =
\frac{ b_{mn}}{a_{mn}}\, \omega_{mn}^2.$$
It is evident from Eq.~(\ref{GamBR}) that, at $|\omega_{mn}| \gg
\gamma_{mn}$, the relaxation rate $\Gamma_{mn}$ is proportional to
the noise spectrum $S_H(\omega_{mn})$,
\ba \label{GBR} \Gamma_{nm}^R = b_{mn}\,S_H(\omega_{mn}),  \ea
with the coefficient $b_{mn} = \sum_{\alpha=1}^N |\langle m |
\sigma_z^\alpha | n \rangle |^2$,  as it should be for the
Bloch-Redfield rate. Transitions between states $\ket{m}$ and
$\ket{n}$ separated by a zero Hamming distance ($a_{mn} = 0$) are
also described by the Redfield rate~(\ref{GBR}).

In the absence of high-frequency noise, with $\eta = 0$ and $S_H =
0$, the function (\ref{GHa}) peaks at zero frequency:
$G_{mn}^H(\omega) = 2 \pi \delta(\omega)$.  In this case the
relaxation rate  (\ref{GamD}) of the many-qubit system is determined
by the Gaussian line shape,
\ba \label{GMarc} \Gamma_{nm}^M = \Delta^2_{mn} \, \sqrt{\frac{2
\pi}{W^2_{mn}} } \exp \left[ - \frac{ (E_m - E_n -
\varepsilon_{mn})^2}{2 W^2_{mn}} \right].  \hspace{0.25cm} \ea
This line shape is typical of the Marcus formulas
\cite{Amin08,Yang02}. The multiqubit tunneling amplitude
$\Delta^2_{kn}(0)$ is determined by the expression
\ba \label{DM}\Delta^2_{mn} \equiv \Delta^2_{mn}(0) = \left( b_{mn}
- \frac{|c_{mn}|^2}{a_{mn}} \right)\, W^2 + \nn
\left| \,T_{mn} - i \langle m | \dot n \rangle - d_{mn}\,
\varepsilon_L - \omega_{mn} \, \frac{c_{mn}}{a_{mn}} \,\right|^2 .
 \ea

\subsection{Relaxation rate of the single qubit}

We assume that the  single qubit is described by a Hamiltonian
(\ref{HamTLS}),
\ba   H_S = -\frac{h}{2} \, \sigma_z - \frac{\Delta}{2} \, \sigma_x,
\nonumber \ea
with a bias $h$, a tunneling amplitude $\Delta$, and energy
splitting $\Omega_0 = \sqrt{\Delta^2 + h^2}.$ The energy basis
$\{\ket{k}\}$ has only two states, $\ket{1}$ and $\ket{2}$. These
states can be found from the equation: $H_S \ket{m} = E_m \ket{m}$.
In Eq.~(\ref{GamD}) for the rate $\Gamma_{nm}$ we assume that $n =
1$ and $m = 2$. The ground state $\ket{n}$ and the first excited
state $\ket{m}$ have the energies: $E_m = - E_n = \Omega_0/2.$ We
work in the energy basis where $T_{mn} = 0.$ For the single qubit we
obtain the following set of parameters,
\ba \label{abS} a_{mn} = 4\,\frac{ h^2}{ \Omega_0^2},\;  b_{mn} =
\frac{\Delta^2}{\Omega_0^2 }, \, c_{mn} =  - 2\,\frac{ h \Delta}{
\Omega_0^2}, \,  d_{mn} =0, \hspace{0.5cm} \ea
so that $B_{mn} = 0$ and $\Delta^2_{mn}(\omega)  = \Delta^2/a_{mn} $
(see sections \ref{Basis} and \ref{ConGL} for definitions). It
follows from Eq.~(\ref{GamD}) that in the case of hybrid noise the
single-qubit relaxation rate combines both, Bloch-Redfield and
Marcus, formulas,
\ba \label{GamS} \Gamma_{nm} = \Delta^2 \, \int
\frac{d\omega}{2\pi}\; \frac{S_H(\omega)}{\omega^2 + \gamma_{mn}^2}
\times \nn \sqrt{\frac{2 \pi}{a_{mn} W^2} }\; \exp \left[ - \frac{
(\Omega_{0} - \omega - a_{mn} \,\varepsilon_L )^2}{2\, a_{mn}\, W^2}
\right], \ea
where $ \gamma_{mn} = a_{mn}\frac{ \eta T}{2}. $ In the limit of
small $\Delta$ the rate (\ref{GamS}) corresponds to the formula
(\ref{GaMR}) shown in Sec.~\ref{SingleQ}.

\section{Dissipative evolution of a 16-qubit system}
\label{Dickson16}

\begin{figure}
\includegraphics[width=0.4\textwidth]{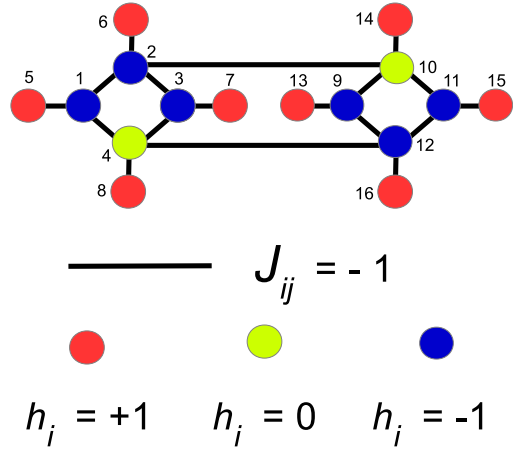}
\caption{\label{DStructure} The 16-qubit instance. Qubits are
denoted as circles, FM couplings  as black lines. Colors correspond
to biases applied to the qubits.}
\end{figure}

In this section we analyze dynamics of the 16-qubit structure
depicted in Fig.~\ref{DStructure}. The structure is determined by
the Dickson instance, which was proposed in Ref.~\cite{Dickson11}
and investigated in details in Ref.~\cite{Dickson13}. The energy
spectrum of the problem features an extremely small gap between the
ground and first excited states. The existence of such a gap
presents a computational bottleneck for quantum annealing. An
experimental technique to overcome this difficulty by individual
tuning qubit's transverse fields has been demonstrated in
Ref.~\cite{King17}. Nevertheless, a theoretical analysis of
dissipative dynamics in this system presents a real challenge.

\begin{figure}
\includegraphics[width=0.5\textwidth]{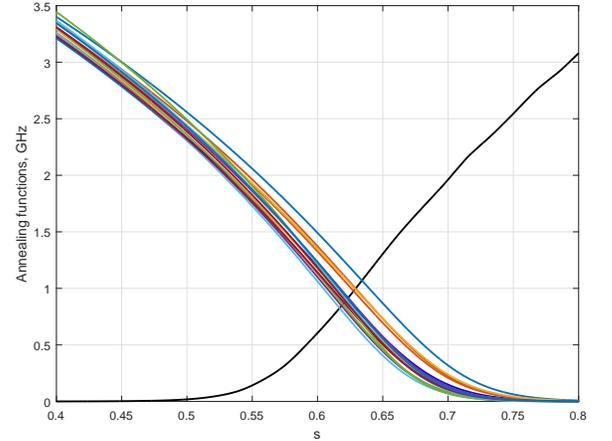}
\caption{\label{AnFun} Annealing parameters ${\cal B}(s)$ (black
line) and tunneling amplitudes $\Delta_1(s), ... \Delta_{16}(s)$
(all other colors)
  plotted as functions of $s$. }
\end{figure}

The probability distribution of the qubits is governed by the master
equation (\ref{rT5a}) with the relaxation matrix given by
Eq.~(\ref{GamD}).
 The qubits are described the Hamiltonian $H_S$
(\ref{HS}). In the problem Hamiltonian $H_P$ (\ref{HDP}) we have
ferromagnetic couplings between qubits, $J_{ij} = -1$, for every
pair of coupled qubits. Two internal qubits have zero biases,
$h_4=h_{10}=0$, whereas the other internal qubits are negatively
biased, with
$$h_1=h_2=h_3=h_9=h_{11}=h_{12}=-1.$$
All external qubits have positive biases:
$$ h_5 = h_6 = h_7 = h_8 = h_{13} = h_{14} = h_{15} = h_{16} = 1.$$
We use the annealing curves $\Delta_\alpha(s) = \Delta_\alpha {\cal
A } (s)$ and ${\cal B}(s)$ plotted in Fig.~\ref{AnFun}. We also take
into account minor variations of the annealing schedule between the
qubits.

\begin{figure}
\includegraphics[width=0.5\textwidth]{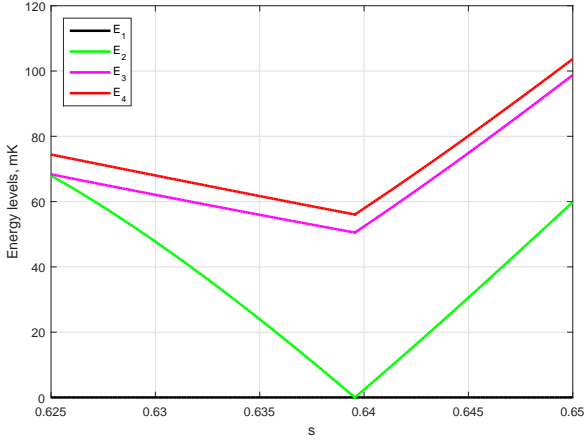}
\caption{\label{EnLev16} Four energy levels of the 16-qubit system
as functions of the annealing parameter near the anticrossing of two
lowest energy levels. Energies are counted from the energy $E_1$ of
the ground state.}
\end{figure}

 The spectrum of the system has an extremely
small energy gap, $E_2-E_1$=0.011~mK, between the ground and the
first excited states \cite{Dickson13}. This gap is located at $ s^*
= 0.6396.$ In Fig.~\ref{EnLev16} we show the four lowest energy
levels of the system near the anticrossing.  The most interesting
annealing dynamics happen in the interval $ s_1 < s < s_2,$ where
$s_1 = 0.625$ and $s_2 = 0.65$.


The two diabatic states with the lowest energies, $\ket{GM}$ and
$\ket{\Sigma}$, are given by the expressions
\ba \label{GM} \ket{\rm GM} = \ket{ \downarrow_1 \downarrow_2
\downarrow_3 \downarrow_4 \downarrow_9 \downarrow_{10}
\downarrow_{11} \downarrow_{12}} \otimes \nn  \ket{ \downarrow_5
\downarrow_6 \downarrow_7 \downarrow_8  \downarrow_{13}
\downarrow_{14} \downarrow_{15} \downarrow_{16} },\;\, \nn
\ket{\Sigma} = \ket{ \uparrow_1 \uparrow_2 \uparrow_3 \uparrow_4
\uparrow_9 \uparrow_{10} \uparrow_{11} \uparrow_{12} } \otimes
\nn\ket{\rightarrow_5 \rightarrow_6 \rightarrow_7 \rightarrow_8
\rightarrow_{13} \rightarrow_{14} \rightarrow_{15} \rightarrow_{16}
}. \ea
Here we introduce the eigenstates $\ket{\uparrow_\alpha}$ and
$\ket{\downarrow_\alpha}$ of the  matrix $\sigma_z^\alpha$, and also
their superposition $\ket{\rightarrow_\alpha}$,
\ba
 \sigma_z^\alpha \ket{\uparrow_\alpha } =
\ket{\uparrow_\alpha },\quad  \sigma_z^\alpha \ket{\downarrow_\alpha
} = - \ket{\downarrow_\alpha },\nn \ket{\rightarrow_\alpha} =
\frac{1}{\sqrt{2}} ( \ket{\uparrow_\alpha } + \ket{\downarrow_\alpha
} ).\quad \; \nonumber \ea
 More details can be found in Ref. \cite{Dickson13} and in
the supplementary information for that paper. It follows from
Fig.~2c of Ref.~\cite{Dickson13} that, before the anticrossing at $
s < s^* $, the instantaneous eigenstates of the 16-qubit system
coincide with the diabatic states: $\ket{1} = \ket{\Sigma},$
$\ket{2} = \ket{\rm GM}. $ After the anticrossing point at $s >
s^*$, we have the reverse situation, with $\ket{1} = \ket{\rm GM}$
and $\ket{2} = \ket{\Sigma}.$ Although the experimental results
provided in Ref.~\cite{Dickson13} were in accordance with the
physical intuition given in the paper, no theoretical analysis was
provided. This was due to the lack of an open quantum theory that
takes into account both low-frequency and high-frequency noises.
Here, we apply our approach to provide a theoretical explanation of
the experimental results of Ref.~\cite{Dickson13}.

The presence of a very small gap and the time-dependence of the
system Hamiltonian, which becomes nonadiabatic near the minimum gap,
make the problem instance in Fig.~\ref{DStructure} difficult to
analyze within one theoretical framework in all regions during the
annealing. As such, some tricks are necessary to choose the proper
basis as we discuss next.

\subsection{Rotation of the basis}

\begin{figure}
\includegraphics[width=0.6\textwidth]{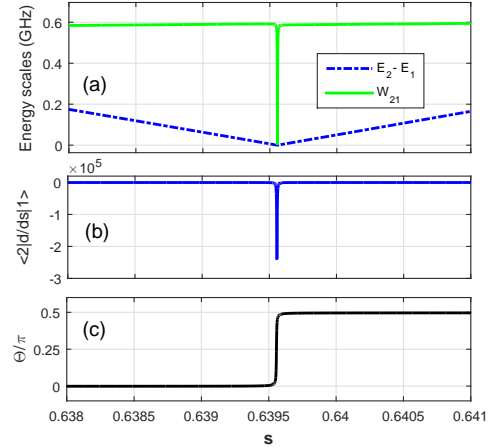}
\caption{\label{GapTheta} (a) The energy scales $E_2-E_1$ and the
line width $ W_{21} = W a_{21}$ calculated in the instantaneous
basis of qubit states. These variables are shown as functions of the
annealing parameter $s$ near the anticrossing point. According to
(\ref{tauC}), the scales $E_2-E_1$ and $W_{21}$ determine the
inverse correlation time $\tau_{21}^{-1}$ of the bath. (b) The
$s$-dependence of the matrix element $\langle 2 | \frac{d}{d s} |
1\rangle$ calculated with Eq. (\ref{MdotN}). This matrix element is
a part of the renormalized tunneling coefficient $\bar T_{mn}$
(\ref{TkC}) and, thus, of the rate $\Gamma_{21}$ (\ref{GamB}). (c)
The optimal rotation angle $\Theta /\pi$ obtained as a solution of
Eq.~(\ref{ThetaS}). }
\end{figure}

The dissipative dynamics of the qubits coupled to a heat bath is
described by the master equations (\ref{rT5a}). These equations are
derived with the proviso that the rate $\Gamma_{nm}$ of the
relaxation  (\ref{GamD}) between the states $\ket{n}$ and $\ket{m}$
is much less than the inverse time scale $\tau_{mn}^{-1}$ given by
Eq.~(\ref{tauC}), so that: $ \Gamma_{nm} \tau_{mn} \ll 1. $ For the
system of 16 qubits under study the perturbation requirement breaks
down at the anticrossing point as it is evident from
Fig.~\ref{GapTheta}a. Here we plot the energy gap, $E_2 - E_1$,
between two instantaneous eigenstates of the Hamiltonian $H_S$ (see
dot-dashed blue line), and also the MRT line width, $W_{21} = a_{21}
W $ (see continuous green line), as functions of the annealing
parameter $s$. At $s = s^*$ both parameters, $E_2-E_1$ and $W_{21}$,
become extremely small, leading to a diverging correlation time
$\tau_{mn}$ (\ref{tauC}). At the same time, $\Gamma_{12}$ becomes
very large due to the contribution from $\tilde{T}_{21}$. Both of
these break the applicability condition (\ref{GamTau}) in the
instantaneous energy basis. Moreover, the time dependence of the
Hamiltonian can create nonzero off-diagonal elements of the density
matrix near the minimum gap due to nonadiabatic transitions. These
terms do not decay quickly as required by our theory. As we shall
see, all these issues can be resolved by rotating the basis. This is
equivalent to the introduction of the pointer basis as described in
Refs.~\cite{Boixo16, Boixo14,Zurek03}.
We rotate the two anticrossing states as:
\ba \label{rot1}\ket{1'} &=& \cos\Theta \;\ket{1} + \sin \Theta
\;\ket{2}, \nn \ket{2'} &=& -\sin\Theta \;\ket{1} + \cos \Theta\;
\ket{2}. \ea
The rotation angle $\Theta$  can depend on the annealing parameter
$s$ and therefore on time $t$. For real eigenstates $\ket{1}$ and
$\ket{2}$ it follows that
\ba \langle 2' | \frac{d}{ds} | 1'\rangle  = \langle 2 |
\frac{d}{ds} | 1\rangle + \frac{d \Theta}{ds}. \nonumber \ea
We choose the rotation angle $\Theta(s)$ such that in the rotated
basis
\ba \label{LZC} \langle 2' | \frac{d}{ds} | 1'\rangle = 0.\ea
 This means that
the $s$-dependence of the angle $\Theta$ is determined by
\ba \label{ThetaS} \frac{d \Theta}{ds} = \frac{\langle 2 | \frac{d
H_S}{ds} | 1\rangle }{E_2 - E_1}. \ea
Notice that  (\ref{LZC}) assures minimum quantum transition between
the two states $\ket{1'}$ and $\ket{2'}$ near the anticrossing and
therefore minimum generation of off-diagonal elements of the density
matrix. As we show in Appendix \ref{ApxRates}, it also resolves all
issues with the applicability condition (\ref{GamTau}) discussed
above.

A solution of Eq.~(\ref{ThetaS}) is shown in Fig.~\ref{GapTheta}c.
In the beginning of annealing $\Theta \simeq 0$, so that the rotated
basis coincides with the instantaneous basis.  Near the anticrossing
point, at $s = s^*$, the angle $\Theta$ rapidly switches to $\pi/2$.
We notice that, with the condition (\ref{LZC}), the renormalized
matrix element $\tilde T_{2'1'}$ (\ref{TmnT}) is
\ba \label{T21} \tilde T_{2'1'} = \langle 2' | H_S | 1'\rangle =
\frac{E_2 - E_1}{2} \, \sin 2 \Theta. \ea
This is zero before ($\Theta = 0$) and after  ($\Theta = \pi/2$) the
anticrossing due to the sine function and is very small at the
anticrossing due to the small gap ($E_2-E_1\approx 0$). This means
that $\tilde T_{2'1'}$ does not contribute to the rate
$\Gamma_{1'2'}$ keeping it small, within the applicability range of
our model. For this example we rotate only the two lowest-energy
states that have an anticrossing. However, the rotation can be
applied to anticrossing excited states as well, if necessary.

\subsection{Thermal enhancement of the success probability}

The goal of annealing is to reach the ground state at the end of the
evolution. It follows from Fig.~2c of Ref.~\cite{Dickson13} that at
the end of annealing, at $t = t_f$, the ground state of the 16-qubit
system coincides with the state $\ket{\rm GM}$ shown in
Eq.~(\ref{GM}). We therefore define the success probability as the
probability $P_{\rm GM}$ to observe the system in $\ket{\rm GM}$ at
$t = t_f$. Figure 3 of Ref.\cite{Dickson13} demonstrates the
temperature dependence of $P_{\rm GM}$. It is clear from this figure
that, at sufficiently fast annealing ($t_f \leq 100$~ms), $P_{\rm
GM}$ grows with increasing temperature from 20 to 40~mK and
decreases after. Our goal is to reproduce this non-monotonic
behavior with our open quantum model. Notice that we do not aim to
precisely fit the experimental data to the results of our model.

We solve numerically the master equations (\ref{rT5a}) with the
relaxation rates given by the convolution formula (\ref{GamD})
written in the rotated basis (\ref{rot1}). We perform simulation for
the total anneal time $t_f$ between 0.04 to 4 ms. An example  of
$P_{\rm GM}$ as a function of time is shown in Fig.~\ref{PGM}b in
Appendix \ref{ApxRates}. Fig.~\ref{PTF} plots the success
probability, $P_{\rm GM}(t_f),$ as a function of temperature $T$ for
different speeds of annealing characterized by the anneal time
$t_f$. In the theoretical calculations we assume that $\eta = 0.1$
and $W = 20$~mK.
\begin{figure}
\includegraphics[width=0.5\textwidth]{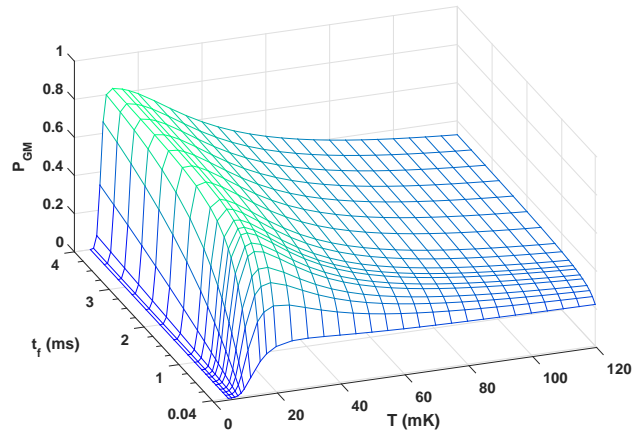}
\caption{\label{PTF} End-of-annealing probability to be in the
$\ket{\rm GM}$ state, $P_{\rm GM}$, as a function of temperature $T$
and the anneal time $t_f$ at $\eta = 0.1$ and $W = 20$~mK.}
\end{figure}
Figure~\ref{PTF} reproduces the results shown in Fig.~3 of
Ref.\cite{Dickson13}, including the enhancement $P_{\rm GM}$ at low
temperatures and its reduction at $T \geq 40$~mK. As mentioned in
Ref.~\cite{Dickson13},  this decrease may be related to the
excitement of the high-energy levels separated from the two lowest
states by a gap of order 40~mK (see the spectrum in
Fig.~\ref{EnLev16}).

\section{Conclusions}

In this paper, we have derived a set of master equations describing
a dissipative evolution of an open quantum system interacting with a
complex environment. The environment has low-frequency and
high-frequency components, as in the case of realistic qubits
affected by the hybrid bath, which includes $1/f$ and Ohmic noise. A
part of the system-bath interaction is treated in a nonperturbative
way. This treatment allows us to combine the Bloch-Redfield and
Marcus approaches to the theory of open quantum systems and obtain
the relaxation rates, which are well-suited for the description of
dissipative dynamics of many-qubit quantum objects, such as quantum
annealers. The relaxation rates are expressed in the convenient
convolution form clearly showing the interplay between the low- and
high-frequency noise. The main results of the paper are given by the
master equations (\ref{rT5a}) with the relaxation rates
(\ref{GamD}). As an illustration, we apply the theory to the
16-qubit quantum annealer investigated in Ref.~\cite{Dickson13}. The
instance studied there features an extremely small gap between the
ground and first excited states. With the proper rotation of the
basis, we have solved the master equations and theoretically
confirmed the main experimental findings of Ref.~\cite{Dickson13}.
The results of the paper may be useful for understanding a
dissipative evolution of various systems, from chromophores in
quantum biology \cite{Yang02,Ghosh11,Lambert13} to qubits in
real-world quantum processors \cite{Gibney17, Mohseni17}.

\acknowledgements

We acknowledge fruitful discussions with Evgeny Andriyash, Mark
Dykman, Andrew King, Chris Rich and Vadim Smelyanskiy. We also thank
Joel Pasvolsky and Fiona Hanington
 for careful reading of the paper.

\appendix

\section{Notations}

\label{ApxNot}

In this appendix we assemble notations used throughout the paper, so
that the main part of the paper becomes easier to follow.
In a chosen basis $\{\ket{n}\}$ the matrix elements of the system
Hamiltonian $H_S$ (\ref{HS}) and those of the Pauli matrix
$\sigma_z^\alpha$ of the $\alpha$-qubit are denoted as
\ba \label{ESmn} E_n = \langle n | H_S | n \rangle, \quad T_{mn} =
\langle m | H_S | n \rangle, \nn \sigma^\alpha_n = \langle n |
\sigma_z^\alpha | n\rangle, \quad \sigma^\alpha_{mn} = \langle m |
\sigma_z^\alpha | n\rangle.
 \ea
For combinations of the matrix elements we introduce the following
notations:
\ba \label{abcd} a_{mn} = \sum_{\alpha} (\sigma_m^\alpha -
\sigma_n^\alpha)^2, \; b_{mn} = \sum_\alpha |\sigma^\alpha_{mn}|^2,
\hspace{0.75cm}
 \\ c_{mn} = \sum_\alpha \sigma^\alpha_{mn} (\sigma_m^\alpha
- \sigma_n^\alpha), \; d_{mn} = \sum_\alpha \sigma^\alpha_{mn}
(\sigma_m^\alpha +\sigma_n^\alpha). \nonumber \ea
%
The parameters of the bath are defined as
\ba \varepsilon_L = \int \frac{d\omega}{2\pi}
\frac{S_L(\omega)}{\omega}, \quad \varepsilon_H = \frac{\eta
\omega_c}{2\pi}, \quad  \varepsilon = \varepsilon_L + \varepsilon_H,
\nn W^2 = \int \frac{d\omega}{2 \pi} S_L(\omega) = 2 \,\varepsilon_L
T, \hspace{1.5cm} \ea
with $\eta$ and $\omega_c$ defined in Sec.~III-E. We also introduce
\ba \varepsilon_{mn} = a_{mn} \varepsilon_L, \quad W_{mn}^2 = a_{mn}
W^2, \nn   \eta_{mn} = a_{mn} \eta, \quad \gamma_{mn} =
\frac{1}{2}\, \eta_{mn} T.  \ea
For a time-dependent basis $\ket{n(t)}$, we write the following
functions of time:
 \ba &&\omega_{mn}(t) = E_m(t) - E_n(t), \nn
&&\phi_{mn}(t) = \int_0^t d\tau\, \omega_{mn}(\tau), \nn &&\tilde
T_{mn} = T_{mn} - i\, \langle m |\dot n\rangle, \nn &&\bar T_{mn} =
T_{mn} - i\, \langle m |\dot n\rangle - d_{mn} \,\varepsilon. \ea
The coefficients $A_{mn}, B_{mn}$ used in Eqs.~(\ref{GamD}) and
(\ref{DeKN}) are defined as
\ba \label{ABmn} A_{mn} =   \bar T_{mn} - \omega_{mn} \,
\frac{c_{mn}}{a_{mn}},
\nn B_{mn} = \frac{a_{mn} b_{mn} - |c_{mn}|^2}{a_{mn}^2} =  \nn
\frac{1}{2 a_{mn}^2} \; \sum_{\alpha \beta} | (\sigma_m^{\alpha} -
\sigma_n^{\alpha} ) \,\sigma_{mn}^{\beta} - (\sigma_m^{\beta} -
\sigma_n^{\beta} ) \,\sigma_{mn}^{\alpha} |^2. \ea

\section{ High-frequency dissipative functions} \label{ApxDF}

For the Ohmic high-frequency bath characterized by the spectrum
(\ref{SH}) the dissipative functions $f_H, g_H$ and a correlator
$K_H = \ddot f_H$ are given by the formulas
\ba \label{fH} &&f_H(t) = \frac{\eta}{2 \pi} \, \ln \left[ ( 1 + i
\omega_c t) \, \frac{\sinh(\pi T t)}{\pi T t} \right],  \;\;\\
&&g_H(t) = \frac{\eta}{2 \pi} \, \frac{\omega_c}{1 + i \omega_c t} -
i \frac{\eta T}{2} \, \left[ \coth (\pi T t) - \frac{1}{\pi T t}
\right], \nn &&K_H(t) = \frac{\eta}{2 \pi}\,\left\{
\left(\frac{\omega_c}{1 + i \omega_c t}\right)^2 - \left[\frac{\pi
T}{\sinh(\pi T t)}\right]^2 + \frac{1}{t^2} \right\},\nonumber  \ea
provided that the cutting frequency $\omega_c$ is much higher than
the temperature, $\hbar \omega_c \gg k_B T.$ We notice that at large
times, $t \gg \frac{\hbar}{\pi k_B T}$, all the three functions
vanish. In Eq.~(\ref{SH}) we introduce $\eta$ as a small
dimensionless coupling constant \cite{Leggett87},  and $\omega_c$ as
a large cutting frequency of the high-frequency noise. For the Ohmic
bath, the spectral density $\chi_H''(\omega)$ is defined as
\ba \label{chiH} \chi_H''(\omega) = \frac{\eta \omega}{2}\,
e^{-|\omega|/\omega_c}.  \ea
We assume independent noise sources coupled to every qubit, each
described by the above-mentioned formulas with identical parameters.

\section{Correlator $ \tilde K_{mn}^{m'n'}(t,t')$ } \label{ApxQC}

Here we calculate the correlator (\ref{KmnA}) of the nondiagonal
bath variables with indexes $m\neq n$ and $m'\neq n'$. In this case,
the bath operators are defined by Eq.~(\ref{QTmn}), and the
correlator  $\tilde K_{mn}^{m'n'}(t,t')$ is given by the formula
\ba \tilde K_{mn}^{m'n'}(t,t') = e^{i\phi_{mn}(t)} e^{i
\phi_{m'n'}(t') } \times \nn \langle {\cal S}^\dag_m(t) [ Q_{mn}(t)
- \tilde T_{mn}(t) ] {\cal S}_n(t) \times \nn {\cal S}^\dag_{m'}(t')
[ Q_{m'n'}(t') - \tilde T_{m'n'}(t') ] {\cal S}_{n'}(t') \rangle.
\ea
This correlator can be represented as a sum of four components:
\ba \label{Cor1}\tilde K_{mn}^{m'n'}(t,t') = e^{i \int_0^t d\tau\,
\omega_{mn}(\tau) + i \int_0^{t'} d \tau \,\omega_{m'n'}(\tau)}
\times \nn \{ (i) + (ii) + (iii) + (iv) \}, \hspace{1cm}  \ea
with
\ba \label{Cor2} &&(i) = \tilde T_{mn}\, \tilde T_{m'n'}\;
F_{mn}^{m'n'}(t,t'),  \\ &&(ii) = -\tilde T_{mn} \langle {\cal
S}_m^\dag(t) {\cal S}_n(t) {\cal S}_{m'}^\dag(t') Q_{m'n'}(t') {\cal
S}_{n'}(t') \rangle, \nn &&(iii) = -\tilde T_{m'n'} \langle {\cal
S}_m^\dag(t) Q_{mn}(t) {\cal S}_n(t) {\cal S}_{m'}^\dag(t') {\cal
S}_{n'}(t') \rangle, \nn &&(iv) = \langle {\cal S}_m^\dag(t)
Q_{mn}(t) {\cal S}_n(t) {\cal S}_{m'}^\dag(t') Q_{m'n'}(t') {\cal
S}_{n'}(t') \rangle, \nonumber \ea
where
\ba \label{Fap2} F_{mn}^{m'n'}(t,t') = \langle {\cal S}_m^\dag(t)
{\cal S}_n(t) {\cal S}_{m'}^\dag(t') {\cal S}_{n'}(t') \rangle.
 \ea
The bath variable $Q_{mn}$ is defined in (\ref{Qmn}), and ${\cal
S}_n$ is the $S$-matrix of the bath given by Eq.~(\ref{snA}).

\subsection{ Term $(i)$ and the functional $F_{mn}^{m'n'}$ }

To calculate the functional $F_{mn}^{m'n'}(t,t')$ (\ref{Fap2}) we
consider a more complicated term
\ba \label{Fap3} {\cal F}_{mn}^{m'n'}(t,\tau; t',\tau') =  \nn
\langle {\cal S}_m^\dag(t,\tau) {\cal S}_n(t,\tau) {\cal
S}_{m'}^\dag(t',\tau') {\cal S}_{n'}(t',\tau') \rangle, \ea
where the modified $S$-matrix of the bath, $  {\cal S}_n(t,\tau),$
is defined by Eq.~(\ref{snB}). We notice that
\ba \label{FmnA} F_{mn}^{m'n'}(t,t') = {\cal F}_{mn}^{m'n'}(t,t;
t',t').  \ea
 The
functional $ {\cal F}_{mn}^{m'n'}(t,\tau; t',\tau')$ obeys two
differential equations:
\ba \label{Fap4} \frac{d}{d \tau} \,{\cal F}_{mn}^{m'n'}(t,\tau;
t',\tau') = i \sum_\alpha \tilde \sigma_{nm}^\alpha(t) \times \nn
\langle {\cal S}_m^\dag(t,\tau) Q_\alpha(\tau) {\cal S}_n(t,\tau)
{\cal S}_{m'}^\dag(t',\tau') {\cal S}_{n'}(t',\tau') \rangle, \nn
\frac{d}{d \tau'} \,{\cal F}_{mn}^{m'n'}(t,\tau; t',\tau') =  i
\sum_\alpha \tilde \sigma_{n'm'}^\alpha(t') \times \nn \langle {\cal
S}_m^\dag(t,\tau) {\cal S}_n(t,\tau)  {\cal S}_{m'}^\dag(t',\tau')
Q_\alpha(\tau') {\cal S}_{n'}(t',\tau') \rangle, \ea
where we introduce the notation $\tilde \sigma_{nm}^\alpha(t) =
\sigma_n^\alpha(t) - \sigma_m^\alpha(t). $ With the Wick theorem
\cite{BSH}, we have to take all possible pairings of the Gaussian
operators $Q_\alpha(\tau)$ and $Q_\alpha(\tau')$ in Eq.~(\ref{Fap4})
with other operators. As a result, we obtain
\ba \label{Fap5} &&\frac{d}{d \tau} \,\ln {\cal
F}_{mn}^{m'n'}(t,\tau; t',\tau') = \sum_\alpha
\tilde\sigma_{mn}^\alpha(t) \times  \nn &&\int_0^\tau d\tau_1
\left[\, \sigma_n^\alpha(t) \,K_\alpha(\tau,\tau_1) -
\sigma_m^\alpha(t) \,K_\alpha(\tau_1,\tau)\, \right] - \nn
&&\sum_\alpha \tilde \sigma_{mn}^\alpha(t) \,\tilde
\sigma_{m'n'}^\alpha(t')\, \int_0^{\tau'} d\tau_1\,  \,
K_\alpha(\tau,\tau_1), \nn
&&\frac{d}{d \tau'} \,\ln {\cal F}_{mn}^{m'n'}(t,\tau; t',\tau') =
\sum_\alpha \tilde\sigma_{m'n'}^\alpha(t') \times \nn
&&\int_0^{\tau'} d\tau_1 \left[\, \sigma_{n'}^\alpha(t')
\,K_\alpha(\tau',\tau_1) - \sigma_{m'}^\alpha(t')
\,K_\alpha(\tau_1,\tau')\, \right] - \nn &&\sum_\alpha \tilde
\sigma_{mn}^\alpha(t) \, \tilde
\sigma_{m'n'}^\alpha(t')\int_0^{\tau} d\tau_1\,  \,
K_\alpha(\tau_1,\tau'). \ea
The solution of these two equations is given by the expression
\ba \label{FmnB} &&\ln {\cal F}_{mn}^{m'n'}(t,\tau; t',\tau') =
 \nn &&- \sum_\alpha \tilde \sigma_{mn}^\alpha(t)\,
\tilde\sigma_{m'n'}^\alpha(t') \int_0^\tau d\tau_1 \int_0^{\tau'}
d\tau_2 \, K_\alpha(\tau_1,\tau_2) + \nn
&&\sum_\alpha \tilde \sigma_{mn}^\alpha(t)\, \int_0^\tau d\tau_1
\int_0^{\tau_1} d\tau_2 \times \nn  &&\left[ \sigma_n^\alpha(t)\,
K_\alpha(\tau_1,\tau_2) - \sigma_m^\alpha(t)\,
K_\alpha(\tau_2,\tau_1) \,\right] + \nn
&&\sum_\alpha \tilde \sigma_{m'n'}^\alpha(t')\, \int_0^{\tau'}
d\tau_1 \int_0^{\tau_1} d\tau_2 \times \nn  &&\left[
\sigma_{n'}^\alpha(t')\, K_\alpha(\tau_1,\tau_2) -
\sigma_{m'}^\alpha(t')\, K_\alpha(\tau_2,\tau_1) \,\right] . \ea
It follows from Eq.~(\ref{FmnA}) that the functional
$F_{mn}^{m'n'}(t,t')$ of the Gaussian bath is described by the
formula
\ba \label{FmnC} F_{mn}^{m'n'}(t, t') =  \nn \exp\left\{ -
\sum_\alpha (\sigma_m^\alpha - \sigma_n^\alpha)(t) \,
(\sigma_{m'}^\alpha - \sigma_{n'}^\alpha)(t')\times \right.\nn
\left. \int_0^t d\tau_1 \int_0^{t'} d\tau_2 \,
K_\alpha(\tau_1,\tau_2) + \right. \nn \left.
\sum_\alpha  (\sigma_m^\alpha - \sigma_n^\alpha)(t)\, \int_0^t
d\tau_1 \int_0^{\tau_1} d\tau_2 \times \right.\nn \left. \left[
\sigma_n^\alpha(t)\, K_\alpha(\tau_1,\tau_2) - \sigma_m^\alpha(t)\,
K_\alpha(\tau_2,\tau_1) \,\right] + \right. \nn \left.
\sum_\alpha (\sigma_{m'}^\alpha - \sigma_{n'}^\alpha)(t')\,
\int_0^{t'} d\tau_1 \int_0^{\tau_1} d\tau_2 \times \right.\nn
\left.\left[ \sigma_{n'}^\alpha(t')\, K_\alpha(\tau_1,\tau_2) -
\sigma_{m'}^\alpha(t')\, K_\alpha(\tau_2,\tau_1) \,\right] \right\}
. \ea
Here $K_\alpha(t,t')$ is the correlation function (\ref{KS}) of the
free bath.

\subsection{ Terms $(ii)$ and $(iii)$ }

Using the Wick theorem \cite{BSH}, we find that the terms $(ii)$ and
$(iii)$ are proportional to the correlators
\ba \langle {\cal S}_m^\dag(t) {\cal S}_n(t) {\cal S}_{m'}^\dag(t')
Q_{m'n'}(t') {\cal S}_{n'}(t') \rangle = \nn  i\, \sum_\alpha
\sigma_{m'n'}^\alpha(t')\, F_{mn}^{m'n'}(t,t') \times \nn \left\{
(\sigma_n^\alpha - \sigma_m^\alpha)(t) \int_0^t dt_1
K_\alpha(t_1,t') + \right.\nn\left. \int_0^{t'} dt_1
[\,\sigma_{n'}^\alpha(t')\, K_\alpha(t',t_1) -
\sigma_{m'}^\alpha(t')\, K_\alpha(t_1,t')  \,] \right\}, \nn
\langle {\cal S}_m^\dag(t) Q_{mn}(t) {\cal S}_n(t) {\cal
S}_{m'}^\dag(t'){\cal S}_{n'}(t') \rangle = \nn  i\, \sum_\alpha
\sigma_{mn}^\alpha(t)\, F_{mn}^{m'n'}(t,t') \times \nn \left\{
(\sigma_{n'}^\alpha - \sigma_{m'}^\alpha)(t') \int_0^{t'} dt_1
K_\alpha(t,t_1) + \right.\nn\left. \int_0^{t} dt_1
[\,\sigma_{n}^\alpha(t)\, K_\alpha(t,t_1) - \sigma_{m}^\alpha(t)\,
K_\alpha(t_1,t)  \,] \right\}. \ea

\subsection{ Term $(iv)$ and the total correlator}

The last term in Eq.~(\ref{Cor2}) can be written as
\ba \langle {\cal S}_m^\dag(t) Q_{mn}(t) {\cal S}_n(t) {\cal
S}_{m'}^\dag(t') Q_{m'n'}(t') {\cal S}_{n'}(t') \rangle = \nn
\sum_\alpha \sigma_{mn}^\alpha(t) \,\sigma_{m'n'}^{\alpha} (t')
\,K_\alpha(t,t') \, F_{mn}^{m'n'}(t,t') + \nn i^2 \sum_{\alpha
\alpha'} \sigma_{mn}^\alpha(t) \,\sigma_{m'n'}^{\alpha'}(t') \,
F_{mn}^{m'n'}(t,t') \times \nn \left\{ \int_0^{t} dt_1
[\,\sigma_{n}^\alpha(t)\, K_\alpha(t,t_1) - \sigma_{m}^\alpha(t)\,
K_\alpha(t_1,t)  \,] + \right. \nn \left. (\sigma_{n'}^\alpha -
\sigma_{m'}^\alpha)(t') \int_0^{t'} dt_1 K_\alpha(t,t_1)
\right\}\times \nn
\left\{ \int_0^{t'} dt_1 [\,\sigma_{n'}^{\alpha'}(t')\,
K_{\alpha'}(t',t_1) - \sigma_{m'}^{\alpha'}(t')\,
K_{\alpha'}(t_1,t') \,] + \right. \nn \left. (\sigma_{n}^{\alpha'} -
\sigma_{m}^{\alpha'})(t) \int_0^{t} dt_1 K_{\alpha'}(t_1,t'))
\right\}. \hspace{1cm} \ea
The correlation function (\ref{Cor1}) of the operators $\tilde
Q_{mn}(t)$ and $\tilde Q_{m'n'}(t')$ has the form
\ba \label{Cor3} \tilde K_{mn}^{m'n'}(t,t')
= e^{i\phi_{mn}(t)} e^{i \phi_{m'n'}(t') } \times  \hspace{0.5cm} \\
\{ {\cal K}_{mn}^{m'n'}(t,t') + {\cal U}_{mn}^{m'n'}(t,t') \,{\cal
V}_{m'n'}^{mn}(t',t) \} \, F_{mn}^{m'n'}(t,t'), \nonumber \ea
where
\ba \label{Cor4} {\cal K}_{mn}^{m'n'}(t,t') =
\sum_\alpha\sigma_{mn}^\alpha(t) \,
\sigma_{m'n'}^\alpha(t') \, K_\alpha(t,t'),  \hspace{0.5cm} \\
{\cal U}_{mn}^{m'n'}(t,t') = \tilde T_{mn}(t) - \nn i \sum_\alpha
\sigma_{mn}^\alpha(t) \left\{(\sigma_{n'}^\alpha -\sigma_{m'}^\alpha
)(t') \int_0^{t'} dt_1 K_\alpha(t,t_1) + \right.\nn \left.\int_0^t
dt_1 [ \,\sigma_n^\alpha(t) \,K_\alpha(t,t_1) - \sigma_m^\alpha(t)\,
K_\alpha(t_1,t)\, ] \right\}, \nn
{\cal V}^{mn}_{m'n'}(t',t) = \tilde T_{m'n'}(t') - \nn i
\sum_{\alpha} \sigma_{m'n'}^{\alpha}(t') \left\{(\sigma_{n}^{\alpha}
-\sigma_{m}^{\alpha})(t) \int_0^{t} dt_1 K_\alpha(t_1,t') +
\right.\nn \left.\int_0^{t'} dt_1 [\, \sigma_{n'}^{\alpha}(t')\,
K_{\alpha}(t',t_1) - \sigma_{m'}^{\alpha}(t')\, K_{\alpha}(t_1,t')
\,] \right\}. \nonumber  \ea
We notice that
\ba \label{DagF} &&\left[ F_{mn}^{m'n'}(t,t') \right]^\dag =
F_{n'm'}^{nm}(t',t),  \nn
&&\left[ \,{\cal K}_{mn}^{m'n'}(t,t') \,\right]^\dag = {\cal
K}_{nm}^{n'm'}(t',t), \nn &&\left[\, {\cal U}_{mn}^{m'n'}(t,t')\,
\right]^\dag = {\cal V}_{nm}^{n'm'}(t,t'), \nn &&\left[ \,{\cal
V}^{mn}_{m'n'}(t',t) \,\right]^\dag = {\cal U}^{nm}_{n'm'}(t',t).
\ea

\subsection{ Correlators and dissipative functions}

The correlation function of the bath (\ref{Cor3}) and the functional
(\ref{FmnC}) can be rewritten in terms of the dissipative functions
defined by Eq.~(\ref{fgA}). Taking into account integrals, such as
\ba \int_0^t dt_1 \int_0^{t'} dt_2\, K_\alpha(t_1,t_2) = f_\alpha(t)
+ f_\alpha^*(t') - f_\alpha(t-t'), \nn \int_0^t dt_1 \int_0^{t_1}
dt_2 K_\alpha(t_1,t_2) = - i\, \varepsilon_\alpha \, t +
f_\alpha(t), \nonumber \hspace{1cm} \ea
we find that
\ba \label{FDa} \ln F_{mn}^{m'n'}(t,t') = \sum_\alpha
(\sigma_m^\alpha - \sigma_n^\alpha) (\sigma_{m'}^\alpha -
\sigma_{n'}^\alpha ) \, f_\alpha(t-t') - \nn i \sum_\alpha
\varepsilon_\alpha [ (\sigma_m^\alpha)^2 - (\sigma_n^\alpha)^2 ]\, t
- i \sum_\alpha \varepsilon_\alpha [ (\sigma_{m'}^\alpha)^2 -
(\sigma_{n'}^\alpha)^2 ]\, t' + \nn \sum_\alpha ( \sigma_m^\alpha -
\sigma_n^\alpha )(\sigma_n^\alpha - \sigma_{m'}^\alpha +
\sigma_{n'}^\alpha ) f_\alpha(t) - \hspace{1cm} \nn
\sum_\alpha ( \sigma_{m'}^\alpha - \sigma_{n'}^\alpha
)(\sigma_{m'}^\alpha + \sigma_{m}^\alpha - \sigma_{n}^\alpha )
f_\alpha^*(t') \hspace{1.25cm} \\
+ \sum_\alpha ( \sigma_{m'}^\alpha - \sigma_{n'}^\alpha )
\sigma_{n'}^\alpha f_\alpha (t') - \sum_\alpha ( \sigma_m^\alpha -
\sigma_n^\alpha ) \sigma_m^\alpha f_\alpha^*(t) .\nonumber  \ea
Here the matrix elements $\sigma_m^\alpha$ and $\sigma_n^\alpha$ are
taken at time $t$, whereas the elements $\sigma_{m'}^\alpha$ and
$\sigma_{n'}^\alpha$ depend on the time $t'$. The total
reorganization energy $\varepsilon_\alpha$ is defined by
Eq.~(\ref{epA}).

With Eq.~(\ref{FDa}) we can calculate the correlator $\langle {\cal
S}_m^\dag(t) {\cal S}_n(t) \rangle $. It follows from
Eqs.~(\ref{snA}) and (\ref{Fap2}) that this correlator  is
determined  by Eqs.~(\ref{FmnA}) and (\ref{FmnC}) where we have to
put $t' = 0$:
\ba \label{smnT} \langle {\cal S}^\dag_m(t) {\cal S}_n(t) \rangle =
F_{mn}^{m'n'}(t, t'=0) =  \;\;\nn \exp \left[ - \sum_\alpha
(\sigma_m^\alpha - \sigma_n^\alpha )^2\, f'_\alpha(t) \right] \,e^{i
\vartheta_{mn}(t) }, \ea
with the phase \ba \vartheta_{mn}(t) = \sum_\alpha\, [
(\sigma_m^\alpha)^2 - (\sigma_n^\alpha)^2 ] \; [ f_\alpha''(t) -
\varepsilon_\alpha t]. \ea

Taking into account Eqs.~(\ref{fL}) and (\ref{fH}) we find that
\ba \label{SmSn}
 \langle {\cal S}^\dag_m(t) {\cal S}_n(t) \rangle = e^{-\frac{1}{2}
 W^2_{mn} t^2 } \, e^{i \vartheta_{mn}(t) } \times \nn
 \left[ \sqrt{ 1 + \omega_c^2 t^2} \; \frac{\sinh (\pi T t) }{\pi T t} \right]^{- \frac{\eta_{mn}}{2 \pi}}.
\ea
Low-frequency,  $W_{mn}^2 = a_{mn} W^2$, and high-frequency,
$\eta_{mn} = a_{mn} \eta$, parameters are proportional to the
Hamming distance $a_{mn}$ between states $\ket{m}$ and $\ket{n}$
(\ref{abc}). At $a_{mn} \neq 0$ the function $\langle {\cal
S}^\dag_m {\cal S}_n \rangle$ rapidly decays within the time scale
of $1/W_{mn}$.  The average value of the operator (\ref{QTmn}) also
does not survive during the annealing process since
\ba \label{Qav} \langle \tilde Q_{mn}(t) \rangle \sim  e^{i
\omega_{mn} t } \, \langle {\cal S}^\dag_m(t) {\cal S}_n(t) \rangle
\sim 0. \ea

\subsection{Selection rules}

The real part of the exponent (\ref{FDa}) has the form
\ba  \label{ReF} \Re \,\{\,\ln F_{mn}^{m'n'}(t,t') \,\} = \hspace{1cm}   \\
\sum_\alpha (\sigma_m^\alpha - \sigma_n^\alpha) (\sigma_{m'}^\alpha
- \sigma_{n'}^\alpha ) \, f_\alpha'(t-t') \hspace{1cm} \nn -
\sum_\alpha [ (\sigma_m^\alpha - \sigma_n^\alpha)^2 - (
\sigma_{m'}^\alpha - \sigma_{n'}^\alpha )^2 ] \, \frac{f_\alpha'(t)
- f_\alpha'(t')}{2} \nn
 - \sum_\alpha (\sigma_m^\alpha - \sigma_n^\alpha + \sigma_{m'}^\alpha
- \sigma_{n'}^\alpha)^2 \, \frac{f_\alpha'(t) + f_\alpha'(t')}{2} .
\nonumber \ea
We expect now that the time interval $t - t'$ is short enough, so
that
 $t \sim t' \gg |t - t'|.$ However, we have $t \sim t_f$ and $t'
 \sim t_f,$ where $t_f$ is the total annealing time.
Notice also that the functions $f_\alpha(t)$ and $f_\alpha(t')$ are
growing with time. It follows from the formulas in
Appendix~\ref{ApxDF} that the high-frequency parts of the functions
$f_\alpha(t)$ and $f_\alpha(t')$ are linearly increasing with time:
$f_H(t) \sim \eta T t$ at $ t \gg \frac{\hbar}{\pi T} \gg
\frac{1}{\omega_c}.$ We assume that environments coupled to
different qubits are described by the same dissipative functions as
we do in Sec.~\ref{HybridNoise}. The low-frequency component
$f_L(t)$ grows with time as well. This means that, during the
annealing run, the contribution of the last line in Eq.~(\ref{ReF})
suppresses the functional $F_{mn}^{m'n'}(t,t')$ if the prefactor
$(\sigma_m^\alpha - \sigma_n^\alpha + \sigma_{m'}^\alpha -
\sigma_{n'}^\alpha)^2$ is not equal to zero. The only surviving term
in the matrix $F_{mn}^{m'n'}(t,t')$ should have the set of indexes
such that the relation
\ba \label{select1} \sigma_m^\alpha - \sigma_n^\alpha +
\sigma_{m'}^\alpha - \sigma_{n'}^\alpha = 0   \ea
is satisfied during the entire annealing process at  $t \sim t' \gg
|t-t'|$. The relation (\ref{select1}) is true at every annealing
point for the indexes:
\ba \label{select2} m = n' , \;\;\; n = m'.  \ea
In this case, the real part of the exponent (\ref{ReF}) depends on
the time interval $t - t'$ only:
\ba  \label{ReFS} \Re \,\{\,\ln F_{mn}^{nm}(t,t') \,\} =  -
\sum_\alpha (\sigma_m^\alpha - \sigma_n^\alpha)^2
 \, f_\alpha'(t-t'). \;\;\ea
We notice that the selection rules (\ref{select2}) are derived
provided that  $\sigma_m^\alpha = \langle m | \sigma_z^\alpha
|m\rangle \neq 0$ at some $m$ and $\alpha$.

At the condition (\ref{select2}), we obtain the following expression
for the functional $F_{mn}(t,t') \equiv F_{mn}^{nm}(t,t')$,
\ba \label{FDb} F_{mn}(t,t') = \exp \left\{ - \sum_\alpha
(\sigma_m^\alpha - \sigma_n^\alpha)^2 f_\alpha(t-t') \,- \right. \nn
\left. i\, \sum_\alpha \varepsilon_\alpha\, [ \,(\sigma_m^\alpha)^2
- (\sigma_n^\alpha)^2 \,]\, (t-t') \,+  \right. \\ \left. 2 i \,
\sum_\alpha ( \sigma_m^\alpha - \sigma_n^\alpha )\, \sigma_m^\alpha
\, [\, f_{\alpha}''(t) - f_\alpha''(t') \, ] \right\}. \nonumber \ea

With the selection rules (\ref{select2}), the correlator
(\ref{Cor3}) takes the form
\ba \label{Cor5} \langle \tilde Q_{mn}(t) \tilde Q_{m'n'}(t')
\rangle = \delta_{m n'} \delta_{n m'} \,\langle \tilde Q_{mn}(t)
\tilde Q_{nm}(t') \rangle, \hspace{0.5cm}  \ea
where
\ba \label{Cor6} \tilde K_{mn}(t,t') \equiv \langle \tilde Q_{mn}(t)
\tilde Q_{nm}(t') \rangle =
e^{i \int_{t'}^t d\tau \omega_{mn}(\tau) } \times  \hspace{1cm}  \\
\left[ \sum_\alpha |\sigma_{mn}^\alpha |^2 \, K_\alpha(t,t') + {\cal
U}_{mn}(t,t')\, {\cal U}^*_{mn}(t',t) \right]  F_{mn}(t,t').
\nonumber \ea
The function  ${\cal U}_{mn}(t,t') \equiv {\cal U}_{mn}^{nm}(t,t')$
is defined in Eq.~(\ref{Cor4}),
\ba \label{UmnA} {\cal U}_{mn}(t,t') = \tilde T_{mn}(t) -  \qquad \\
i\, \sum_\alpha \sigma_{mn}^\alpha(t)  (\sigma_m^\alpha -
\sigma_n^\alpha ) (t')\, \int_0^{t'} dt_1 K_\alpha(t,t_1) - \nn i\,
\sum_\alpha \sigma_{mn}^\alpha(t)  \int_0^t dt_1 [
\,\sigma_n^\alpha(t) \, K_\alpha (t,t_1) - \sigma_m^\alpha(t)
K_\alpha (t_1,t) \, ]. \nonumber \ea
Taking into account the integrals, such as
\ba \int_0^{t'} dt_1 K_\alpha(t,t_1) =  i \,[ \, g_\alpha(t) -
 g_\alpha(t-t') \,], \ea
we find the function  ${\cal U}_{mn}(t,t')$,
\ba \label{UmnB} {\cal U}_{mn}(t,t') = \tilde T_{mn} - \sum_\alpha
\sigma_{mn}^\alpha (\sigma_m^\alpha - \sigma_n^\alpha ) \, \tilde
g_\alpha(t-t') - \nn \sum_\alpha \sigma_{mn}^\alpha \,
\sigma_m^\alpha \, [\, \tilde g_\alpha(t) + \tilde g_\alpha(-t) \,].
  \hspace{0.5cm}\ea
The function $\tilde g_\alpha(t)$ is defined as: $\tilde g_\alpha(t)
= g_\alpha(t) - \varepsilon_\alpha,$ where $g_\alpha$ is shown in
(\ref{fgA}).
 We also notice that
\ba \tilde g_\alpha(t) + \tilde g_\alpha(-t) = - 2\, \int
\frac{d\omega}{2 \pi} \, \chi''_\alpha(\omega)\, \frac{ 1 - \cos
\omega t}{\omega}. \ea
Taking into account that the dissipative function $g_\alpha(t)$ goes
to zero at $ t \rightarrow \infty$, we obtain the steady-state
expression for the function ${\cal U}_{mn}(t,t')$,
\ba \label{UmnC} {\cal U}_{mn}(t,t') = \bar T_{mn}  - \sum_\alpha
\sigma_{mn}^\alpha (\sigma_m^\alpha - \sigma_n^\alpha ) \,
g_\alpha(t-t'), \hspace{0.5cm} \ea
where
\ba \label{TmnB} \bar T_{mn} = T_{mn} - i\, \langle m |\dot n
\rangle - \sum_\alpha \sigma_{mn}^\alpha (\sigma_m^\alpha +
\sigma_n^\alpha )\, \varepsilon_\alpha. \ea
We presume that all matrix elements of qubit operators are taken at
 time $t$.

For the case of qubits coupled to environments described in
Sec.~\ref{HybridNoise}, we obtain the following expression for the
bath correlator (\ref{Cor6}):
\ba \label{Cor7a} \tilde K_{mn}(t,t')  = e^{i \omega_{mn}(t) \tau}
\, e^{-a_{mn} f(\tau)} \, e^{i (\zeta_n - \zeta_m) \tau } \times \nn
\{ b_{mn} \ddot f(\tau) + [ \bar T_{mn} - c_{mn} g(\tau) ] \,[ \bar
T_{mn}^* - c_{mn}^* g(\tau) ] \}, \ea
where $\tau = t - t'.$ The parameter $\zeta_n$ has a meaning of a
polaron shift \cite{Boixo16, Boixo14},
\ba \label{zetan} \zeta_n = \sum_\alpha \varepsilon_\alpha
(\sigma_n^\alpha)^2.  \ea
We notice that the polaron shift (\ref{zetan}) contains
contributions of both, low-frequency, $\varepsilon_L$, and
high-frequency, $\varepsilon_H$, parts of the bath reorganization
energy since $\varepsilon_\alpha = \varepsilon_L + \varepsilon_H. $
The polaron shift introduced in Eqs.~(5) and (6) of
Ref.~\cite{Boixo16} depends on low-frequency noise only.

The bath correlator (\ref{Cor7a}) is characterized by a short
correlation time $\tau_{mn}$. The parameter $\tau_{mn}^{-1}$  can be
evaluated as
\ba \label{tCor} \frac{1}{\tau_{mn}} = {\rm max} \{ |E_m - E_n|,
W_{mn} \}. \ea

\subsection{ Cross-correlator $\langle \tilde Q_{mn}(t)\, \tilde
Q_{kk}(t') \rangle $ } \label{CrossQ}

The cross-correlations are characterized by the function
\ba \label{Cor7} \langle \tilde Q_{mn}(t)\, \tilde Q_{kk}(t')
\rangle = -\, e^{i\, \phi_{mn}(t)} \sum_{\alpha'} \dot
\sigma_k^{\alpha'}(t') \times \nn \int_0^{t'}  d\tau' \langle
\,{\cal S}_m^\dag(t) \,[\, Q_{mn}(t) - \tilde T_{mn} \,]\, {\cal
S}_n(t) \times \nn {\cal S}_k^\dag(t',\tau')\, Q_{\alpha'}(\tau') \,
{\cal S}_k(t',\tau')\, \rangle .   \ea
Taking into account the definitions of the $S$-matrix (\ref{snA})
and (\ref{snB}) and applying the Wick theorem \cite{BSH} for the
Gaussian operators of the bath,
 we find that the correlator (\ref{Cor7}) is proportional
to the function $\langle {\cal S}_m^\dag(t) {\cal S}_n(t) \rangle$
given by Eq.~(\ref{smnT}),
\ba \label{Cor8} \langle \tilde Q_{mn}(t)\, \tilde Q_{kk}(t')
\rangle = {\cal J}_{mn}^k(t,t')\times \langle {\cal S}_m^\dag(t)
{\cal S}_n(t) \rangle, \ea
where
\ba {\cal J}_{mn}^k(t,t') = i\, {\cal Z}_{mn}(t) \; e^{
i\phi_{mn}(t) } \sum_{\alpha} \dot \sigma_k^{\alpha}(t') \times
\hspace{0.5cm} \nn \int_0^{t'} d\tau  \left\{ \sigma_k^{\alpha}(t')
 \int_0^{\tau} d\tau_1 [ K_{\alpha}(\tau,\tau_1) -
K_{\alpha}(\tau_1, \tau) ] \, + \right. \nn \left.
(\sigma_n^{\alpha} - \sigma_m^{\alpha})(t) \, \int_0^t d\tau_1\,
K_{\alpha}(\tau_1, \tau) \right\} - \nn
  e^{ i\phi_{mn}(t) }\, \sum_\alpha \sigma_{mn}^\alpha(t)\, \dot
\sigma_k^{\alpha}(t') \int_0^{t'} d\tau K_\alpha(t,\tau),
\hspace{0.5cm} \ea
with the renormalized tunneling coefficient $ {\cal Z}_{mn}(t)$
defined as
\ba {\cal Z}_{mn}(t) = \tilde T_{mn} - i \sum_\alpha
\sigma_{mn}^\alpha(t) \times \nn  \int_0^t d t_1 \,[\,
\sigma_n^\alpha(t)\, K_\alpha(t,t_1) - \sigma_m^\alpha(t) \,
K_\alpha(t_1,t)\, ]\,. \ea
The most important fact here is that the cross-correlator
(\ref{Cor8}) is proportional to the function $\langle {\cal
S}_m^\dag(t) {\cal S}_n(t) \rangle$, which rapidly decays during the
annealing process according to Eq.~(\ref{smnT}). Therefore, the
cross-correlators (\ref{Cor7}) between diagonal and off-diagonal
operators of the bath give no contribution to the evolution equation
(\ref{rT1}).

\section{Derivation of master equations}
\label{ApxMeq}

The time evolution of the system operators $\langle
\Lambda_{nn}\rangle $ is governed by Eq.~(\ref{rT1}). We transform
this relation to the set of master equations for the probabilities
$P_n$ defined by Eq.~(\ref{Pnb}). To do that, we have to calculate
products of bath and system operators, such as $ \langle Q^I_{mn} \,
\Lambda_{mn}\rangle$, where $\langle \ldots \rangle $ means
averaging over free bath fluctuations. Operators $\Lambda_{mn}$ and
$Q^I_{mn}$ are given by Eqs.~(\ref{rhoTa}) and (\ref{QTa}).

\subsection{Calculation of the correlator $\langle Q_{mn}^I
\, \Lambda_{mn} \rangle.$ }  \label{ApxQR}

We begin by calculating a more general correlation function
 $\langle Q_{mn}^I
\, \Lambda_{kl} \rangle$ using a perturbation expansion up to the
second order in the bath operators $\tilde Q_{mn}$ (\ref{QTmn}) and
resorting to the methods outlined in \cite{ES81, Ghosh11,
Klyatskin}.
 It follows from Eqs.~(\ref{rhoTa}) and
(\ref{QTa}) that
\ba \label{QR1} \langle Q_{mn}^I(t) \, \Lambda_{kl}(t) \rangle =
\langle U_I^\dag(t)\, \tilde Q_{mn}(t) \, \ket{k(t)}\bra{l(t)}\,
U_I(t) \rangle,  \ea
where the unitary matrix $U_I$ (\ref{UI}) is determined by the
Hamiltonian $H_I$ given by Eq.~(\ref{HIc}). Using functional
derivatives, up to the second order in operators $\tilde Q_{mn}$, we
obtain
\ba \langle  Q_{mn}^I(t) \, \Lambda_{kl}(t) \rangle = \langle
U_I^\dag(t)\, \tilde Q_{mn}(t) \, \ket{k(t)}\bra{l(t)}\, U_I(t)
\rangle = \nn \int dt'\, \tilde K_{m'n'}^{mn}(t',t) \, \left<
\frac{\delta U_I^\dag(t)}{\delta \tilde Q_{m'n'}(t') } \,
\ket{k(t)}\bra{l(t)} \, U_I(t) \right> + \nn
\int dt' \, \tilde K_{mn}^{m'n'}(t,t') \, \left< U_I^\dag(t)  \,
\ket{k(t)}\bra{l(t)} \,\frac{\delta U_I(t)}{\delta \tilde
Q_{m'n'}(t')}\right>, \nonumber \ea
where  $\tilde K_{mn}^{m'n'}$ is the bath correlation function
defined by Eqs.~(\ref{KmnA}), (\ref{KmnB}),  and (\ref{Cor7a}). The
cross-correlation functions, such as $\langle \tilde Q_{mn}(t)
\tilde Q_{n'n'}(t')\rangle$, do not appear in the correlator
$\langle  Q_{mn}^I \, \Lambda_{kl} \rangle $ since they do not
survive the long annealing run. The average value of the bath
operator $\langle \tilde Q_{mn}(t) \rangle$ also gives no
contribution to Eq.~(\ref{QR1}) as it follows from Eqs.~(\ref{SmSn})
and (\ref{Qav}) obtained in Appendix~\ref{ApxQC}.

We see from Eq.~(\ref{UI}) that
\ba \frac{\delta U_I(t)}{\delta \tilde Q_{m'n'}(t_1)} = - i {\cal T}
\left\{ \int_0^t d\tau \frac{\delta H_I(\tau)}{\delta \tilde
Q_{m'n'}(t_1) } e^{- i \int_0^t dt_2 H_I(t_2) } \right\}. \nonumber
\ea
With the Hamiltonian $H_I$ given by Eq.~(\ref{HIc}), we find
\ba \frac{ \delta H_I(\tau)}{\delta \tilde Q_{m'n'}(t_1) } = -
\delta(\tau - t_1)\, \ket{m'(t_1)}\bra{n'(t_1)}. \ea
For the functional derivatives of the unitary matrices $U_I$ and
$U_I^\dag$, we derive the relations:
\ba \frac{\delta U_I(t) }{\delta \tilde Q_{m'n'}(t_1) } = i \,
\theta(t-t_1) \, U_I(t) \, \Lambda_{m'n'}(t_1), \nn
\frac{\delta U_I^\dag(t) }{\delta \tilde Q_{m'n'}(t_1) } = - i \,
\theta(t-t_1) \, \Lambda_{m'n'}(t_1)\, U_I^\dag(t) . \ea
 With these formulas in
mind, we obtain the following expression for the correlator
(\ref{QR1}):
\ba \label{QR2}
 \langle Q_{mn}^I(t) \, \Lambda_{kl}(t)
\rangle = \hspace{1cm} \\ i \int_0^t dt_1 \tilde
K_{mn}^{m'n'}(t,t_1) \, \langle \Lambda_{kl}(t)  \Lambda_{m'n'}(t_1)
\rangle - \nn i \int_0^t dt_1 \tilde K_{m'n'}^{mn}(t_1,t) \, \langle
\Lambda_{m'n'}(t_1) \Lambda_{kl}(t) \rangle . \hspace{0.25cm}
\nonumber \ea
In Eq.~(\ref{rT1}) we have terms such as $\langle Q_{mn}^I
\Lambda_{mn}\rangle.$ Using the selection rules (\ref{Cor5}) we
obtain
\ba \label{QR3}
 \langle Q_{mn}^I(t) \, \Lambda_{mn}(t)
\rangle = \hspace{1cm} \\ i \int_0^t dt_1 \tilde K_{mn}(t,t_1) \,
\langle \Lambda_{kl}(t)  \Lambda_{nm}(t_1) \rangle - \nn i \int_0^t
dt_1 \tilde K_{nm}(t_1,t) \, \langle \Lambda_{nm}(t_1)
\Lambda_{nm}(t) \rangle . \hspace{0.25cm} \nonumber \ea
After the first step,  the evolution equation (\ref{rT1}) turns into
the form
\ba \label{rTT} \frac{d}{dt} \, \langle \Lambda_{nn}\rangle  =
\sum_{m\neq n} \int_0^t dt_1 \tilde K_{mn}(t,t_1) \, \langle
\Lambda_{mn}(t)\Lambda_{nm}(t_1) \rangle - \nn
\sum_{m\neq n} \int_0^t dt_1 \tilde K_{nm}(t_1,t) \, \langle
\Lambda_{nm}(t_1)\Lambda_{mn}(t) \rangle + \{\rm h.c. \},
\hspace{0.5cm} \ea
where $ \{\rm h.c.\}$ is the Hermitian conjugate of the previous
terms. It is of interest that the time evolution of the system
operator $\langle \Lambda_{nn}\rangle $ depends on the behavior of
the correlators, such as $\langle \Lambda_{mn}(t)\Lambda_{nm}(t_1)
\rangle$.

\subsection{Correlator $\langle \Lambda_{mn}(t) \Lambda_{nm}(t') \rangle$ }

As the next step in the derivation of master equations, we calculate
the correlator of system operators in the right-hand side of
Eq.~(\ref{rTT}). Of interest is when the moments of time $t$ and
$t'$ are separated by the short time interval $\tau_{mn}$ such that
$$t - t' \sim \tau_{mn} \ll t. $$ For Eq.~(\ref{rTT}) the parameter
$\tau_{mn}$ corresponds to the correlation time given by
(\ref{tCor}). It follows from (\ref{rTa}) that the evolution of the
operator $\Lambda_{mn}$ is quite slow. The rate of this evolution is
determined by the bath operators, such as $Q^I_{km}$ and $ Q^I_{nk}$
, which are proportional to the off-diagonal elements of qubit Pauli
matrices, $\sigma^\alpha_{km}$ and $ \sigma^\alpha_{nk}$, and also
to the off-diagonal terms such as $\tilde T_{km}$ and $\tilde
T_{nk}$; see Eqs.~(\ref{ETmn}, \ref{sAmn}, \ref{QTmn}, \ref{TmnT})
for definitions. At first glance, this fact allows us to ignore the
variation of $\Lambda_{nm}(t')$ in time during the interval
$\tau_{mn}$. In this case, the correlator of system operators can be
easily calculated:
\ba \label{LaLam} &&\langle  \Lambda_{mn}(t) \Lambda_{nm}(t')
\rangle \simeq \langle \Lambda_{mn}(t) \Lambda_{nm}(t) \rangle = \nn
&&\langle U_I^\dag \ket{m}\langle n|n\rangle \bra{m} U_I\rangle  =
\langle \Lambda_{mm}(t)\rangle.  \ea
We notice, however, that the sum of the off-diagonal elements, such
as
\ba \label{sumSig} \sum_{k\neq n} \sigma^\alpha_{nk}
\,\sigma^\alpha_{kn} =
  1 - (\sigma_n^\alpha)^2,  \ea
is not small, even though each of the components of this sum is
small by itself. Therefore, the system correlators in (\ref{rTT})
should be calculated more precisely.

To do this, we notice that the correlators in question satisfy the
equation
\ba \label{LCorA} i \frac{d}{dt} \langle \Lambda_{mn}(t)
\Lambda_{nm}(t') \rangle = \sum_k \langle Q^I_{km}(t)
\Lambda_{kn}(t) \Lambda_{nm}(t')\rangle -  \nn \sum_k \langle
Q^I_{nk}(t) \Lambda_{mk}(t) \Lambda_{nm}(t')\rangle , \quad \ea
which follows from (\ref{rTa}). We show that Markovian fluctuations
of the bath characterized by zero-frequency susceptibility
$\chi_\alpha(0)$ contribute to the right-hand side of
Eq.~(\ref{LCorA}). This contribution is significant because it is
proportional to the sums $\sum_{k\neq n} |\sigma_{nk}^\alpha|^2$
given by Eq.~(\ref{sumSig}). We choose the system basis where the
off-diagonal elements, such as $\sigma_{nk}^\alpha$,  are small,
whereas the diagonal elements $\sigma^\alpha_n$ can take any values
from the interval $[-1, 1].$ The components of the Pauli matrix
$\sigma_z^\alpha$ are defined by Eq.~(\ref{sAmn}). Notice that the
Markovian contribution to the right-hand side of Eq.~(\ref{LCorA})
can be traced without resorting to the perturbation theory in the
system-bath interaction. In the process, we drop perturbative terms,
which are proportional to individual matrix elements of the Pauli
matrices and also to the off-diagonal elements $\tilde T_{mn}$ of
the system Hamiltonian $H_S$. For this reason, in the bath operators
$Q^I_{mn}$ involved in (\ref{LCorA}) and defined by Eq.~(\ref{QTa}),
we assume that
$$ \tilde Q_{mn}(t) =  e^{i \phi_{mn} }  \sum_\alpha \sigma^\alpha_{mn} \,{\cal S}_m^\dag(t) Q_\alpha(t) {\cal S}_n(t). $$
The first term in the right-hand side of Eq.~(\ref{LCorA}),
\ba \label{LCorB} \langle Q^I_{km}(t) \Lambda_{kn}(t)
\Lambda_{nm}(t')\rangle = \sum_\alpha \sigma_{km}^\alpha  e^{i
\phi_{km} }  \Xi_{mnk}^\alpha(t,t') ,\nonumber  \ea
is proportional to the factor
 \ba \label{LCorC}
\Xi_{mnk}^\alpha(t,t') = \nn \langle U_I^\dag(t) {\cal S}_k^\dag(t)
Q_\alpha(t) {\cal S}_m(t) \, \ket{k}\bra{n} U_I(t) \,
\Lambda_{nm}(t') \rangle. \ea
From here on, matrix elements, such as $\sigma_{km}^\alpha$, and
system operators, such as $\ket{k}\bra{n}$, are taken at time $t$.
According to the Wick theorem \cite{BSH}, in Eq.~(\ref{LCorC}) we
pair the Gaussian bath operator $Q_\alpha(t)$ with the other
operators, which contain bath variables. Notice that, as follows
from the definition in (\ref{snA}), pairings of $Q_\alpha(t)$ in
(\ref{LCorC}) with the matrices ${\cal S}_k^\dag(t)$ and ${\cal
S}_m(t)$ give rise to diagonal matrix elements, such as
$\sigma_k^\alpha$ and $\sigma_m^\alpha$, and, therefore, to products
such as $\sigma_{km}^\alpha \sigma_k^\alpha$ and $\sigma_{km}^\alpha
\sigma_m^\alpha.$ These products do not combine into sums similar to
(\ref{sumSig}). Therefore, this kind of pairing should be omitted.
The system operator $\Lambda_{nm}(t')$ also contains free bath
operators such as $Q_\alpha(t'')$ where $ t'' \leq t' < t.$ The
relation $t' < t$ means that $\Lambda_{mn}(t')$ cannot depend on the
Markovian operator $Q_\alpha(t)$ taken at the future moment of time
$t$. For this reason, we do not consider pairings between
$Q_\alpha(t)$ and $\Lambda_{mn}(t')$ in (\ref{LCorC}).

Taking pairings of $Q_\alpha$ and the operators $U_I^\dag$ and
$U_I$, we obtain
\ba \label{XiA} \Xi_{mnk}^\alpha(t,t') = \hspace{1cm}  \\ \int dt_1
K_\alpha(t_1,t) \left< \frac{\delta U_I^\dag(t)}{\delta
Q_\alpha(t_1) } {\cal S}_k^\dag {\cal S}_m\ket{k}\bra{n} U_I(t)
\Lambda_{nm}(t') \right> + \nn
\int dt_1 K_\alpha(t,t_1) \left< {U_I^\dag (t) \cal S}_k^\dag {\cal
S}_m \ket{k}\bra{n} \frac{\delta U_I^\dag(t)}{\delta Q_\alpha(t_1) }
\Lambda_{nm}(t') \right>.\nonumber \ea
where $K_\alpha(t,t_1) = \langle Q_\alpha(t) Q_\alpha(t_1) \rangle$
is the correlation function of the free bath.  The matrices ${\cal
S}_k^\dag $ and ${\cal S}_m$ are taken at time $t$.
For the functional derivative of the matrix $U_I$ (\ref{UIa}), we
obtain
\ba \label{UIb} \frac{\delta U_I(t)}{\delta Q_\alpha(t_1)} = - i
\,{\cal T} \left\{ \int_0^t d\tau \frac{\delta H_I(\tau)}{\delta
Q_\alpha(t_1) } \;e^{-i \int_0^t dt_2 H_I(t_2) } \right\}. \ea
In the Hamiltonian $H_I(\tau)$ we keep the component that is
proportional to the free bath operator $Q_\alpha(\tau)$,
\ba H_I(\tau) = - \sum_\alpha \sum_{k'\neq l'}
\sigma^\alpha_{k'l'}(\tau) \, e^{i \phi_{k'l'}(\tau)}\times \nn
{\cal S}_{k'}^\dag(\tau) \,Q_\alpha(\tau) \,{\cal S}_{l'}(\tau)
\ket{k'(\tau)}\bra{l'(\tau)}.  \ea
The dominant term in the functional derivative of the Hamiltonian
$H_I$ has the form
\ba
 \frac{\delta H_I(\tau)}{\delta Q_\alpha(t_1)
}  = - \delta(\tau - t_1)\times \hspace{1cm} \\ \sum_{k'\neq l'}
\sigma^\alpha_{k'l'}(t_1) \, e^{i \phi_{k'l'}(t_1)}\, {\cal
S}_{k'}^\dag(t_1)  \,{\cal S}_{l'}(t_1) \ket{k'(t_1)}\bra{l'(t_1)}.
\nonumber \ea
For the derivative (\ref{UIb}), we obtain
\ba \label{dUa} \frac{\delta U_I(t)}{\delta Q_\alpha(t_1)} = i \,
\theta(t-t_1) \times \\ \sum_{k'\neq l'} \sigma^\alpha_{k'l'}(t_1)
\, e^{i \phi_{k'l'}(t_1)}\, U_I(t)\, {\cal S}_{k'l'}(t_1)
\,\Lambda_{k'l'}(t_1), \nonumber \ea
where we introduce the new bath operator
\ba \label{SKL} {\cal S}_{kl}(t) = U_I^\dag(t)\, {\cal
S}_{k}^\dag(t)\, {\cal S}_{l}(t)\, U_I(t). \ea
We also have the following formula
\ba \label{dUb}  \frac{\delta U_I^\dag(t)}{\delta Q_\alpha(t_1)} = -
i \, \theta(t-t_1) \times \\ \sum_{k'\neq l'}
\sigma^\alpha_{k'l'}(t_1) \, e^{i
\phi_{k'l'}(t_1)}\,\Lambda_{k'l'}(t_1)\,  {\cal
S}_{k'l'}(t_1)\,U_I^\dag(t). \nonumber \ea
With Eqs.~(\ref{dUa}) and (\ref{dUb}), the function
$\Xi_{mnk}^\alpha(t,t') $ (\ref{XiA}) takes the form
\ba \label{XiB} \Xi_{mnk}^\alpha(t,t')  = i \sum_{k' l'} \int_0^t
dt_1 K_\alpha(t,t_1) \sigma_{k'l'}^\alpha(t_1) e^{i
\phi_{k'l'}(t_1)} \times  \nn \langle \Lambda_{kn}(t){\cal
S}_{km}(t) {\cal S}_{k'l'}(t_1)  \Lambda_{k'l'}(t_1)
\Lambda_{nm}(t') \rangle   - \quad\nn
 i \sum_{k' l'} \int_0^t dt_1 K_\alpha(t_1,t)
\sigma_{k'l'}^\alpha(t_1) e^{i \phi_{k'l'}(t_1)}\times \nn \langle
\Lambda_{k'l'}(t_1) {\cal S}_{k'l'}(t_1){\cal S}_{km}(t)
\Lambda_{kn}(t) \Lambda_{nm}(t') \rangle. \qquad   \ea
The bath correlator $K_\alpha(t,t_1) = K_\alpha(t-t_1)$ is defined
by Eq.~(\ref{KS}). The Markovian part of this correlator has a sharp
peak at $t=t_1.$ Its contribution to the function
$\Xi_{mnk}^\alpha(t,t')$ is described by Eq.~(\ref{XiB}) where in
all functions of $t_1$ we have to put $t_1 = t$ and, after that,
remove these functions from the integrals over time. We notice that
operators $\Lambda_{kn}(t)$ (\ref{rhoTa}) and ${\cal S}_{k'l'}(t)$
(\ref{SKL}) taken at the same moment of time commute at any sets of
indexes. This fact allows us to calculate the same-time products of
the system operators with relations outlined in (\ref{LaLam}). In
particular, we have
\ba  \Lambda_{kn}(t) \Lambda_{k'l'}(t) &=& \delta_{nk'} \,
\Lambda_{k l'}(t), \nn \Lambda_{k'l'}(t) \Lambda_{kn}(t) &=&
\delta_{kl'} \Lambda_{k'n}(t). \nonumber \ea
The function (\ref{XiB}) now turns to the form
\ba \label{XiC} \Xi_{mnk}^\alpha(t,t')  = i  \int_0^t dt_1
K_\alpha(t,t_1) \times \nn \sum_{l\neq n } \sigma_{nl}^\alpha(t)
e^{i \phi_{nl}(t)} \langle {\cal S}_{km}(t) {\cal S}_{nl}(t)
\Lambda_{kl}(t) \Lambda_{nm}(t') \rangle - \nn
 i \int_0^t dt_1 K_\alpha(t_1,t) \times \nn
\sum_{l\neq k}  \sigma_{lk}^\alpha(t) e^{i \phi_{lk}(t)} \langle
{\cal S}_{lk}(t){\cal S}_{km}(t)   \Lambda_{ln}(t) \Lambda_{nm}(t')
\rangle. \quad \ea
This function appears in (\ref{LCorB}) and, after that, in
Eq.~(\ref{LCorA}) for the correlator $\langle \Lambda_{mn}(t)
\Lambda_{nm}(t') \rangle$. The secular part of
$\Xi_{mnk}^\alpha(t,t') $ should contain the correlator of system
operators with the same set of indexes. In the first term in the
right-hand side of Eq.~(\ref{XiC}) we have to put $k=m$ and $l = n$,
which is impossible since  $l\neq n.$ Therefore, the first component
of $\Xi_{mnk}^\alpha(t,t')$ has no secular term. In the second part
of (\ref{XiC}) we take that $l = m$. This is accepted if $m \neq k.$
Thus, the secular part of the function $\Xi_{mnk}^\alpha(t,t')$ can
be written as
 \ba \label{XiD} \Xi_{mnk}^\alpha(t,t')  = - i
\int_0^t dt_1 K_\alpha(t_1,t) \times \nn  \sigma_{mk}^\alpha(t) e^{i
\phi_{mk}(t)} \langle \Lambda_{mn}(t) \Lambda_{nm}(t') \rangle.
\quad \ea
Here we take into account that $ {\cal S}_{mk}(t){\cal S}_{km}(t) =
1.$ The secular part of (\ref{LCorB}) has the form
\ba \langle Q^I_{km}(t) \Lambda_{kn}(t) \Lambda_{nm}(t')\rangle =
\qquad \\ - i \langle \Lambda_{mn}(t) \Lambda_{nm}(t') \rangle
  \sum_\alpha |\sigma^\alpha_{mk}(t)|^2 \int_0^t dt_1
K_\alpha(t_1,t). \nonumber \ea
 With the same approach, we obtain a secular component of the last term in (\ref{LCorA}):
\ba \langle Q^I_{nk}(t) \Lambda_{mk}(t) \Lambda_{nm}(t')\rangle =
\qquad \\  i \langle \Lambda_{mn}(t) \Lambda_{nm}(t') \rangle
  \sum_\alpha |\sigma^\alpha_{nk}(t)|^2 \int_0^t dt_1
K_\alpha(t,t_1). \nonumber \ea
We notice that nonsecular terms in the right-hand side of
Eq.~(\ref{LCorA}) are proportional to the products of ${\cal S}$
matrices of the bath, such as given by Eq.~(\ref{SmSn}). These terms
rapidly disappear over time.

The contribution of Markovian fluctuations of the bath to the
evolution of the correlator of system operators is described by the
following equation:
\ba \label{LCorD} \frac{d}{dt} \langle \Lambda_{mn}(t)
\Lambda_{nm}(t') \rangle = \qquad \\ - \sum_\alpha  \sum_{k\neq n}
|\sigma^\alpha_{nk}(t)|^2\int_0^t dt_1 K_\alpha(t,t_1)  \langle
\Lambda_{mn}(t) \Lambda_{nm}(t') \rangle + \nn  \sum_\alpha
\sum_{k\neq m}  |\sigma^\alpha_{mk}(t)|^2\int_0^t dt_1
K_\alpha(t_1,t)  \langle \Lambda_{mn}(t) \Lambda_{nm}(t') \rangle .
\; \nonumber \ea
The correlator $K_\alpha(t)$ of the free bath is defined in Section
\ref{SecBath} together with a susceptibility $\chi_\alpha(\omega)$
and the spectrum $S_\alpha(\omega).$  The upper limit $t$ in
(\ref{LCorD}) can be replaced by infinity since the annealing time
scale $t$ is much longer than the correlation time of the bath.
Taking into account the fluctuation-dissipation theorem (\ref{FDT})
and the fact that $\chi_\alpha(0) = 2 \varepsilon_\alpha$, we obtain
\ba \label{KaF} \int_0^\infty dt_1 K_\alpha(t,t_1) =
T\Upsilon_\alpha - i \varepsilon_\alpha, \ea
where $$ \Upsilon_\alpha = \frac{\chi_\alpha''(\omega)}{\omega
}_{|\omega=0} $$ is a negligibly small parameter, $\Upsilon_\alpha =
\Upsilon_H = \eta/2. $ Finally, keeping the main contributions to
(\ref{LCorA}), we derive a simple equation which governs the
short-time evolution of the system correlator:
\ba \frac{d}{dt} \langle \Lambda_{mn}(t) \Lambda_{nm}(t') \rangle =
\qquad\\ i \, \sum_\alpha \varepsilon_\alpha\left( \sum_{k\neq n}
|\sigma_{nk}^\alpha |^2 - \sum_{m\neq k}  |\sigma_{mk}^\alpha |^2
\right) \langle \Lambda_{mn}(t) \Lambda_{nm}(t') \rangle. \nonumber
\ea
The solution of this equation,
\ba \label{LaCorT} \langle \Lambda_{mn}(t) \Lambda_{nm}(t') \rangle
= e^{i (\zeta_m - \zeta_n) (t - t')} \, \langle
\Lambda_{mm}(t')\rangle, \ea
has the additional factor oscillating in time with the frequency
$\zeta_m - \zeta_n$. Here we take into account Eq.~(\ref{sumSig})
and also the definition (\ref{zetan}) of the polaron shift
$\zeta_n$.

\subsection{Equations for system operators $\langle \Lambda_{nn}\rangle$}

As the next step, we substitute the correlation functions
(\ref{LaCorT}) to Eq.~(\ref{rTT}) taking into account that $t' =
t_1$. In the process, the system operator $\langle \Lambda_{mm}(t_1)
\rangle$ is replaced by $\langle \Lambda_{mm}(t) \rangle$  since
this variable is practically unchanged during the correlation time
$\tau_{mn}$ of the function $e^{i \omega_{mn} \tau}\,\tilde
K_{mn}(\tau)$, where $\tau = t - t_1.$  Equation~(\ref{rTT})
transforms to the equation for the system operator $\langle
\Lambda_{nn}\rangle$,
\ba \label{rT5}  \langle \dot \Lambda_{nn} \rangle + \Gamma_n
\langle \Lambda_{nn} \rangle = \sum_m \Gamma_{nm} \, \langle
\Lambda_{mm} \rangle, \ea
with the relaxation matrix $\Gamma_{nm}$,
\ba \label{GamAA} \Gamma_{nm} =   \int_0^\infty d\tau \,\tilde
K_{mn}(\tau) \,e^{i (\zeta_m - \zeta_n ) \tau} + \{\rm h.c.\}, \ea
and  with the rate $\Gamma_n = \sum_m \Gamma_{mn}. $  Notice that
$\tilde K^\dag_{mn}(\tau) = \tilde K_{mn}(-\tau)$, and that the
matrix $\Gamma_{nm}$ has no diagonal elements. The matrix elements
of the qubit operators involved in Eq.~(\ref{GamAA}) are taken at
 time $t$. The bath correlator $\tilde K_{mn}(\tau)$ is
given by Eq.~(\ref{Cor6}), and also by the simpler
expression~(\ref{Cor7a}). Taking these formulas into account, we
derive Eq.~(\ref{GamB}) for the relaxation matrix $\Gamma_{nm}$. It
is of interest that the polaron shifts in Eq.~(\ref{GamAA})
precisely cancel the polaron shifts in the bath correlator
(\ref{Cor7a}) so that the rate (\ref{GamB}) does not contain
$\zeta_n$ and $\zeta_m$. This is especially important for the
multiqubit system where the relative shift of two levels, $\zeta_m -
\zeta_n$, can be quite large. The master equation (\ref{rT5a}) for
the probability distribution of the system over basis states follows
from Eq.~(\ref{rT5}) if we apply the procedure (\ref{Pnb}) that
turns the system operator $\langle \Lambda_{nn} \rangle$ into the
probability $P_n.$

\subsection{Equations for system operators $\langle \Lambda_{mn}\rangle$}

The time evolution of the qubit operator $\langle
\Lambda_{mn}\rangle $, where $m \neq n$, can be found from
Eq.~(\ref{rTa}) averaged over free bath fluctuations. Using the
results of the previous subsection, and also a secular
approximation, we obtain the simple equation for the function
$\langle \Lambda_{mn}\rangle$,
\ba \label{LMN} \langle \dot \Lambda_{mn} \rangle + \left(
\frac{\Gamma_n + \Gamma_m}{2} + i \frac{\delta_n - \delta_m}{2}
\right) \, \langle \Lambda_{mn} \rangle = 0. \ea
Here, as in the previous subsection, the line width of the $n$-level
is defined as $\Gamma_n = \sum_k \Gamma_{kn}$, where the rates
$\Gamma_{kn}$ are determined by Eq.~(\ref{GamAA}). These rates are
given by Eq.~(\ref{GamTime}) for the case of identical environments,
with the spectra $S_\alpha(\omega) = S(\omega).$ In the more general
case different qubits are coupled to different environments, and
these environments have different spectral functions, such as
$S_\alpha(\omega) \neq S_{\beta}(\omega)$ at $\alpha \neq \beta$. In
this case the rate $\Gamma_{nm}$ can be expressed in terms of the
function ${\cal G}_{nm}(\omega)$:
\ba \label{gamG}
 \Gamma_{nm} = {\cal
G}_{nm}(\omega_{mn}).\ea
where
\ba {\cal G}_{nm}(\omega) = \int_{-\infty}^{\infty} d \tau\;
e^{i\omega \tau} \, e^{- \sum_\alpha (\sigma_m^\alpha -
\sigma_n^\alpha)^2 f_\alpha(\tau)} \times \nn \left\{ \sum_\alpha
|\sigma_{mn}^\alpha|^2 \ddot f_\alpha(\tau) + \right. \nn \left.
[\bar T_{mn} - \sum_\alpha \sigma_{mn}^\alpha (\sigma_{m}^\alpha -
\sigma_n^\alpha) g_\alpha(\tau) ] \times\right. \nn \left.  [\bar
T_{mn}^* - \sum_\alpha \sigma_{nm}^\alpha (\sigma_{m}^\alpha -
\sigma_n^\alpha) g_\alpha(\tau) ] \right\}. \ea
The renormalized matrix element $\bar T_{mn}$ is given by
Eq.~(\ref{TmnB}). In accordance with the dispersion relations, the
frequency shift $\delta_n$ is also defined in terms of the function
${\cal G}(\omega)$,
\ba \delta_n = - \sum_k \int \frac{d\omega}{\pi} \, \frac{{\cal
G}_{kn}(\omega)}{\omega - \omega_{kn} }. \ea
Here, there is no simple convolution form for the function ${\cal
G}_{nm}(\omega) $ and for the rate $\Gamma_{nm}$, as it takes place
for the case of identical environments described in Sec.~\ref{BRM}.
We notice that Eq.~(\ref{gamG}) is equivalent to Eq.~(\ref{GamAA})
from the previous subsection.

\section{ Gaussian and Lorentzian line shapes}
\label{ApxGL}

 In this appendix we describe properties of the
functions $G^L_{mn}(\omega)$ and $G^H_{mn}(\omega)$ used in
Sec.~\ref{ConGL}. The low-frequency envelope (with an index $\mu =
L$) and the high-frequency function (with $\mu = H$) are defined as
\ba \label{GLH1} G^{\mu}_{mn}(\omega) = \int_{-\infty}^{\infty}\,
d\tau \, e^{i\omega \tau} \; e^{-a_{mn} f_{\mu}(\tau)}.  \ea
The low-frequency dissipative function $f_L$ is given by
Eq.~(\ref{fL}). The function $G^L_{mn}(\omega)$ is described by a
Gaussian lineshape,
\ba \label{GL1} G^L_{mn}(\omega) = \int d\tau \;e^{i(\omega -
\varepsilon_{mn})\tau - W^2_{mn}\tau^2/2} = \nn \sqrt{\frac{2
\pi}{W^2_{mn}} }\, \exp \left[ - \frac{(\omega - \varepsilon_{mn}
)^2}{ 2 W^2_{mn}} \right], \ea
where  $\varepsilon_{mn} = a_{mn} \varepsilon_L,$ and $W^2_{mn} =
a_{mn} W^2 =  2 \,\varepsilon_{mn}\, T.$ At $\varepsilon_{mn}
\rightarrow 0, $ we have $G^L_{mn}(\omega) \rightarrow 2\pi
\delta(\omega).$

We assume that the high-frequency function $G^H_{mn}(\omega)$,
defined by Eq.~(\ref{GLH1}) where $\mu = H$, includes the case of
strong interaction of qubits with a Markovian heat bath. This bath
is characterized by the flat spectrum $S_H(\omega) = S_H(0)$ and by
the function $f_H(\tau) = \frac{1}{2}\, S_H(0)\, |\tau|$. The
function $G_{mn}^H(\omega)$ (\ref{GLH1}) related to the Markovian
bath has a Lorentzian shape \cite{Amin08},
\ba \label{GMarkov}
 G_{mn}^H(\omega) = \frac{2 \gamma_{mn}}{\omega^2 +
\gamma_{mn}^2}. \ea
In the Markovian case the linewidth $\gamma_{mn}$,
\ba \label{gamn} \gamma_{mn} = \frac{a_{mn}}{2} \, S_H(0), \ea
should be much smaller than the inverse correlation time of the
high-frequency bath. We also expect that the function
$G^H_{mn}(\omega)$ covers a weak interaction of qubits with a
non-Markovian bath characterized by the frequency-dependent spectrum
$S_H(\omega)$. Taking into account that in the weak-coupling limit
the function $f_H$ is small, we expand the exponent in
Eq.~(\ref{GLH1}) and keep linear terms in power of $f_H$.  We obtain
the following formula for the function $G^H_{mn}(\omega)$:
\ba \label{GH1} &&G^H_{mn}(\omega) = \nn &&2 \pi \delta(\omega)
\left[ 1 - a_{mn} \int \frac{d x}{2 \pi} \, \frac{S_H(x)}{x^2} \,
\right] + a_{mn} \frac{S_H(\omega)}{\omega^2}. \hspace{1cm}\ea
Equations (\ref{GMarkov}) and (\ref{GH1}) can be combined by a
straightforward modification of the Markovian expression
(\ref{GMarkov}),
\ba \label{GH2}
 G_{mn}^H(\omega) = \frac{a_{mn}\, S_H(\omega)}{\omega^2 +
\gamma_{mn}^2}. \ea
Evidently, at $S_H(\omega) = S_H(0)$,  Eq.~(\ref{GH2}) includes the
Markovian case (\ref{GMarkov}). In the limit of zero qubit-bath
coupling (at $S_H \rightarrow 0$) both expressions, (\ref{GH1}) and
(\ref{GH2}), turn into $\delta(\omega)$ function, $G_{mn}^H(\omega)
\rightarrow 2\pi \delta(\omega).$ Finally, at nonzero frequencies,
$\omega \gg \gamma_{mn}$, both functions, (\ref{GH1}) and
(\ref{GH2}), are inversely proportional to the frequency squared and
linearly proportional to the noise spectrum $S_H(\omega),$ as it
takes place in the Bloch-Redfield limit: $G_{mn}^H(\omega) = a_{mn}
S_H(\omega)/\omega^2. $ This means that the Lorentzian line shape
(\ref{GH2}) provides an appropriate description of the function
$G_{mn}^H(\omega)$ in the whole range of frequencies.

Both functions, $G^L_{mn}(\omega)$ (\ref{GL1}) and
$G^H_{mn}(\omega)$ (\ref{GH2}), meet the equilibrium condition,
\ba
 \frac{G^{\mu}_{mn}(\omega)}{G^{\mu}_{mn}(-\omega)} = e^{\omega/T},
 \ea
 and the normalization condition,
 \ba
\int \frac{d\omega}{2\pi}\, G^{\mu}_{mn}(\omega) = 1. \ea
The normalization condition directly  follows from the definition
(\ref{GLH1}). We recall that the index $\mu$ takes two values: $\mu
= L, H.$

\section{ Rates and probabilities for the 16-qubit system}
\label{ApxRates}

\begin{figure}
\includegraphics[width=0.8\textwidth]{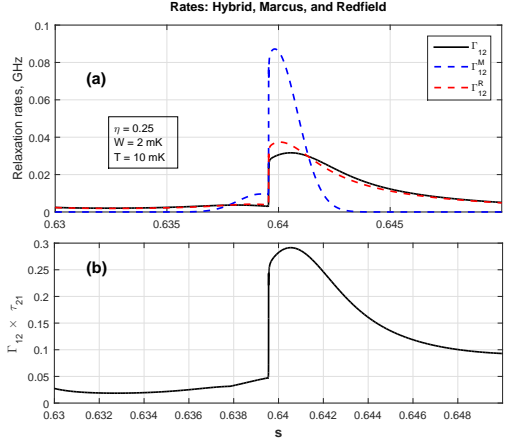}
\caption{\label{GamTa} (a) At large coupling to the HF bath ($\eta =
0.25$) and at a small interaction with the LF noise ($W = $ 2 mK),
the hybrid rate $\Gamma_{12}$ is close to the Bloch-Redfield rate
$\Gamma_{12}^R$. (b) The perturbation parameter $\Gamma_{12}\,
\tau_{21} $ is less than 0.3.}
\end{figure}

\begin{figure}
\includegraphics[width=0.8\textwidth]{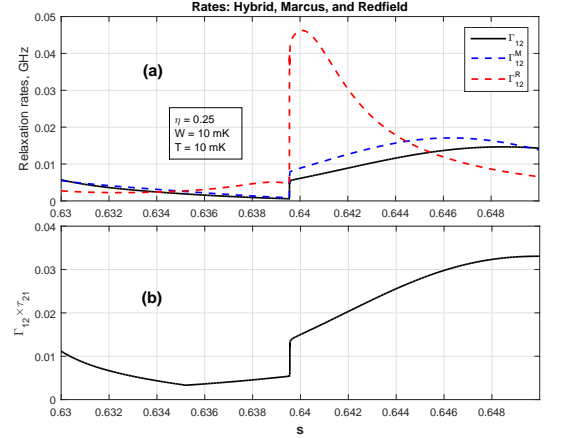}
\caption{\label{GamTb} (a) When both couplings are large ($\eta =
0.25$ and $W = 10$ mK), the hybrid rate $\Gamma_{12}$, taken along
the annealing path, deviates from Bloch-Redfield, $\Gamma_{12}^R$,
and from Marcus, $\Gamma_{12}^M,$ rates. (b) The perturbation
parameter is small, $\Gamma_{12}\,\tau_{21} < 0.04.$ }
\end{figure}

\begin{figure}
\includegraphics[width=0.8\textwidth]{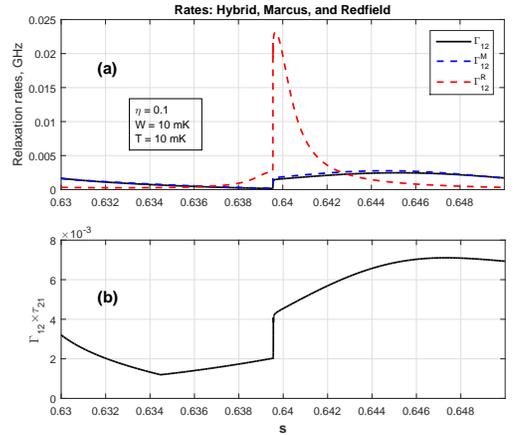}
\caption{\label{GamTc} (a) At sufficiently large coupling of the
system to the low-frequency bath, where $W = 10$ mK, and at small
coupling to the high-frequency bath, with $\eta = 0.1$, the hybrid
rate $\Gamma_{12}$ is close to the Marcus expression
$\Gamma_{12}^M.$  (b) The perturbation parameter is decreasing in
this case: $\Gamma_{12}\, \tau_{21} < 0.01.$}
\end{figure}

In this appendix we calculate the rate $\Gamma_{1'2'}$ (\ref{GamD})
of the system relaxation between the states $\ket{1'}$ and
$\ket{2'}$. These states are defined as superpositions (\ref{rot1})
of the instantaneous ground and first excited states. The rotation
angle $\Theta(s)$ is chosen as the solution of Eq.~(\ref{ThetaS}).
Hereafter we drop primes from the state number and use $\Gamma_{12}$
instead of $\Gamma_{1'2'}$. In Figs.~\ref{GamTa}a, \ref{GamTb}a, and
\ref{GamTc}a we show the $s$-dependence of the hybrid relaxation
rate $\Gamma_{12}$ (black line) in comparison to the Marcus rate
(blue dashed line) and to the Bloch-Redfield rate (red dashed line).
The Bloch-Redfield rate is calculated with Eq.~(\ref{GamBR}),
whereas the Marcus rate is given by Eq.~(\ref{GMarc}). To verify our
approach, in Figs.~\ref{GamTa}b, \ref{GamTb}b, and \ref{GamTc}b we
show the evolution of the supposed-to-be small parameter
$\Gamma_{12}\times \tau_{12}$ (\ref{GamTau}) during the annealing
process. Recall that all rates and parameters are written in the
rotated basis (\ref{rot1}). We keep the same temperature, $T =
10$~mK, in every figure.  We change, however, the coupling constant
$\eta$ and the MRT line width $W$, thus changing a relative
contribution of the high-frequency bath and the low-frequency
environment to the hybrid rate $\Gamma_{12}$. Figure~\ref{GamTa}a is
related to case of the large coupling to the high-frequency bath,
with $\eta = 0.25$, whereas the role of the low-frequency noise is
diminished, with $W = 2$~mK. In this case the hybrid rate
$\Gamma_{12}$ is close to the Bloch-Redfield rate $\Gamma_{12}^R.$
The perturbation parameter remains low during the annealing process,
$\Gamma_{12} \tau_{21} \leq 0.3. $ In Fig.~\ref{GamTb} we consider
the intermediate case where qubit couplings to both low-frequency
and high-frequency environments are quite large, so that $\eta
=0.25$ and $ W = 10$~mK. Here, the hybrid rate $\Gamma_{12}$ differs
from the Redfield rate $\Gamma_{12}^R$ and from the Marcus rate
$\Gamma_{12}^M$. The perturbation parameter is decreasing:
$\Gamma_{12} \tau_{21} < 0.04.$ Fig.~\ref{GamTc} shows that, at the
smaller coupling to the high-frequency bath, where $\eta = 0.1$, and
at the quite strong interaction of qubits with the low-frequency
noise, with $ W = 10$~mK, the hybrid rate $\Gamma_{12}$ almost
coincides with the Marcus rate $\Gamma_{12}^M.$ The validity of
these results is verified by the small parameter $ \Gamma_{12}
\tau_{21} < 0.008 $ shown in Fig.~\ref{GamTc}b.

\begin{figure}
\includegraphics[width=0.5\textwidth]{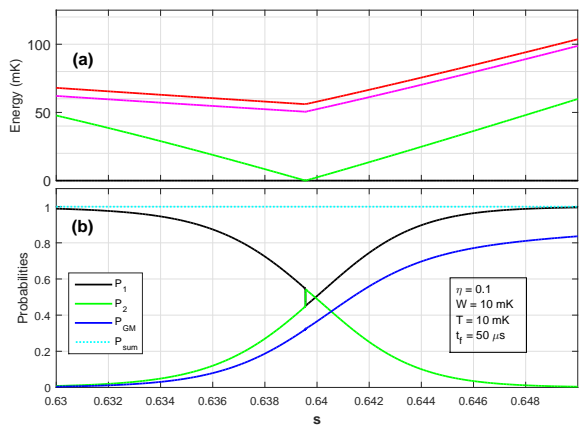}
\caption{\label{PGM} (a) Instantaneous energy spectrum of the
16-qubit system near the anticrossing point. (b) The evolution of
the probabilities $P_1, P_2$ to find the system in the instantaneous
energy eigenstates during the annealing process at the total anneal
time $t_f = 2$ ms and at $\eta = 0.1, W = 10$~mK$, T = 10$~mK. We
also show the time evolution of the probability  $P_{\rm GM}$ for
the system to be in the $\ket{GM}$.}
\end{figure}

In Fig.~\ref{PGM}, in parallel with the energy spectrum,  we plot a
time dependence of the probabilities $P_1$ and $P_2$ to find the
16-qubit system in the instantaneous eigenstates $\ket{1}$ and
$\ket{2}$ of the Hamiltonian $H_S$ (\ref{HS}).
To calculate these probabilities, we obtain the numerical solution
of the master equation (\ref{rT5a}) for the probabilities $P_{1'}$
and $P_{2'}$ to observe the system in the states $\ket{1'}$ and
$\ket{2'}$ (\ref{rot1}). After that, we rotate the basis back, to
the instantaneous energy eigenstates $\ket{1}$ and $\ket{2}.$ The
contribution of the off-diagonal elements of the density matrix to
the probabilities $P_1$ and $P_2$ rapidly disappears as it follows
from Eqs.~(\ref{rNM}) and (\ref{SmSn}). We also show the evolution
of the probability $P_{\rm GM}$ (blue curve) to find the system in
the state $\ket{\rm GM}$ defined in (\ref{GM}). Here, we have a
qualitative agreement with the experimental results shown in Fig.~2d
of Ref.~\cite{Dickson13}).

\end{document}